\documentclass[prc,floatfix,]{revtex4}
\usepackage{graphicx,subfigure}
\usepackage{comment}
\usepackage{amsmath, amsthm, amssymb}
\usepackage{bm}
\def\nuc#1#2{\relax\ifmmode{}^{#1}{\protect{#2}}\else${}^{#1}$#2\fi}

\newcommand {\la} {\langle}
\newcommand {\ra} {\rangle}
\newcommand {\beq} {\begin{eqnarray}}
\newcommand {\eeq} {\end{eqnarray}}
\newcommand {\eeqn} [1] {\label{#1} \end{eqnarray}}%
\newcommand {\eol} {\nonumber \\}
\newcommand {\ve} [1] {\mbox{\boldmath $#1$}}

\begin{document}
\graphicspath{{figures/}}

\title{Three-body model of the $d+A$ system in an antisymmetrised, translationally invariant many nucleon theory}
 \author{R.C. Johnson}
 \affiliation{
  Department of Physics,
 Faculty of Engineering and Physical Sciences,
 University of Surrey,
 Guildford, Surrey GU2 7XH, United Kingdom}

\date{\today}

\begin{abstract}
This paper concerns the theory of direct $A(d,p)B$ reactions and how practical 3-body models of them are related to the optical model potentials describing the interaction of the deuterons constituents and the target $A$. A new definition of the nucleon optical potential is introduced which necessitates a careful examination of the boundary conditions involved in the associated coupled channel formalism. The new definition turns out to be very convenient for making a connection with the theory of the $A(d,p)B$ reaction.  The analysis is carried out within an antisymmetrised, translationally invariant formalism and provides a firm basis from a many-body point of view for the three-body force effects calculated in  [Physical Review C 99 064612(2019)]. 
\end{abstract}

\maketitle
\section{Introduction}\label{A1}
 Recent work\cite{Johnson17}, \cite{Johnson19} has shown how the nucleon optical potential operator for the nucleon$+A$ system, where $A$ denotes an $A$-nucleon target, can be  consistently defined within an antisymmetrised, translationally invariant many nucleon theory that displays explicitly the connection with the underlying nucleon dynamics. In this paper analogous issues for the  $n+p+A$ system are addressed with applications to 3-body models of $A(d,p)B$ reactions.The starting point is the exact $A+2$ scattering wavefunction corresponding to a deuteron incident on an $A$-nucleon target in its ground state. The projection of this wavefunction on to a state of $A$ nucleons in their ground state defines a function with the degrees of freedom of 3 bodies, namely two nucleons and a third body with the mass and degrees of freedom of $A$ nucleons in their ground state. The relation between the effective potential that generates this projection and a new definition of the nucleon optical potential operator is discussed and it is shown explicitly that physical effects arise that are  not taken into account in standard three-body models based on a Hamiltonian with phenomenological nucleon optical potentials. This analysis provides a firm basis from a many-body point of view for the three-body force effects calculated in \cite{Din19} and derived from a non-antisymmetrised many-body model.
 
       In Section \ref{nucleonopt} a new definition of the nucleon optical potential is introduced using the methods of \cite{Johnson19}. Similar ideas are applied to the $n+p+A$ system in Section \ref{dAscattering} and a three-body model Hamiltonian is introduced. In Section \ref{BdA} the physical meaning of the the interaction terms in the three-body Hamiltonian  are discussed in detail.  Section \ref{multscatt} dicusses how the effective interaction in the 3-body model derived in the previous Section can, when appropriate approximations are made, be related to the nucleon optical model operators introduced in Section \ref{nucleonopt}. Conclusions are summarised in Section \ref{conclude}. Some of the detailed algebra involved in the derivations of the main text are included in Appendices \ref{A} to \ref{F}.
\section{Definition of the nucleon optical potential in terms of overlap functions.}\label{nucleonopt}
In this section it is shown how a nucleon optical potential can be defined within an antisymmetrised, translationally invariant many-body theory using the concept of overlap functions. In order to keep the formalism within reasonable bounds of complexity it will be assumed that all Coulomb forces are screened at large distances,  that the $n-p$ mass difference can be neglected and that only two-nucleon interactions are present in the many-nucleon Hamiltonian. None of these assumptions introduce unsurmountable difficulties. The modifications needed when 3-nucleon forces are included are indicated where relevant. In general, the notations of \cite{Johnson17} and \cite{Johnson19}, and more recently \cite{Birse2020}, will be used.
   
A set of one-body overlap function associated with $\mid \Psi\ra\ra $, an N-nucleon state in Fock space, is defined by the expression
\beq \la\la \psi_n \mid a_i\mid \Psi \ra\ra, \label{overlap}\eeq
where the $\mid \psi_n\ra\ra$ are a complete set of $(N-1)$-body states and $a_i$ destroys a nucleon in a one-body state $\mid i\ra$ with quantum numbers $i$.
Knowledge of a set of these overlap functions is sufficient to determine many properties of the state $\mid \Psi\ra\ra$. Note that the states $ a^\dagger_i\mid\psi_n \ra\ra $ are not orthonormal so that the determination of $\mid \Psi\ra\ra $ itself in terms of them requires the inversion of a matrix of their overlaps. Ways of dealing with this step in the context of nucleon optical model theory are discussed in detail in \cite{Birse2020}. However, as emphasised in \cite{Johnson19}, for example, the optical potential is not unique. The definition used here does not require the inversion of overlap matrix. It will be shown that this step is not necessary for a complete discussion of an optical model operator and the on-shell elastic transition operator, although the price to be paid is that the complete $N+A$ scattering wavefunction is not available directly. The new definition, which differs from the definitions used in \cite{Johnson19}, turns out to be very convenient for the discussion of the $n+p+A$ system.   

 A Fock-space scattering state $\mid \Psi_{E,\ve{k}_0}^\epsilon\ra\ra$ relevant to the scattering of a nucleon by an $A$-nucleon target nucleus in its ground state is defined by 
\beq(E-H+\imath \epsilon)\mid \Psi_{E,\ve{k}_0}^\epsilon\ra\ra =\imath \epsilon(2\pi)^{3/2}a^\dagger_{\ve{k}_0}\mid -\ve{k}_0,\psi_0 \ra\ra. \label{Psi1scatt2}\eeq 
where $H$ is the many-nucleon Hamiltonian operator in Fock-space \cite{Johnson17}.

In the limit $\epsilon \rightarrow 0^+$ the ket $\mid \Psi_{E,\ve{k}_0}^\epsilon\ra\ra$ describes an antisymmetric  $(A+1)$-nucleon scattering state of total momentum zero in the overall c.m. system. The incident channel  has an incident nucleon with momentum $\ve{k}_0$. In this channel the $A$-nucleon  target has a total momentum $-\ve{k}_0$ and is in its intrinsic ground state $\psi_0$. All other channel components of  $\mid \Psi_{E,\ve{k}_0}^\epsilon\ra\ra$ have purely outgoing waves asymptotically. The factor $(2\pi)^{3/2}$ arises because $a^\dagger_{\ve{k}_0}$ creates a normalised plane wave state, whereas scattering states are conventionally normalised to an incoming plane wave of unit amplitude. Explicit references to spin and isospin degrees of freedom are omitted. 

 By taking the inner product of both sides of eq.(\ref{Psi1scatt2}) with one of the states $a^\dagger_{\ve{k}}\mid -\ve{k},\psi_n \ra\ra$, with  $n,\ve{k}, $ arbitrary, a set of inhomogeneous coupled equations can be derived for the overlap functions  
$\la\la  -\ve{k},\psi_n\mid a_{\ve{k}}\mid\Psi_{E,\ve{k}_0}^\epsilon\ra\ra $. For a translation invariant $H$ it is convenient to write these equations in terms of the components of the overlap functions in an $A$-nucleon basis, $\mid \Psi(n,\ve{x}=0\ra\ra$, in which  the target is in an intrinsic state $n$ and the target centre-of-mass has a definite position $\ve{x}=0$. These basis states are defined in eqs.(5)-(8) of \cite{Johnson19}.  Using the techniques of \cite{Villars1967} as used in \cite{Johnson17} and \cite{Johnson19} it is found that the overlap components satisfy
 \beq \int d \ve{k}_1\sum_{n'}((E+\imath \epsilon-E_n-\frac{(A+1)}{A}\epsilon_k)\delta(\ve{k}-\ve{k}_1)\delta_{n,n'}-\la\ve{k},n  \mid\hat{ \mathcal{U}}\mid \ve{k}_1,n'  \ra)\la\la\Psi(n',\ve{x}=0 )\mid a_{\ve{k}_1}\mid \Psi_{\ve{k}_0,0}^{(\epsilon)}\ra\ra \eol
=\imath \epsilon (2\pi)^{3/2}\hat{\mathcal{K}}_{n,\ve{k};0,\ve{k}_0}. \label{scattwf4}\eeq
where $E_n$ is the intrinsic energy of the $A$-nucleon zero total momentum state $n$ and the one-nucleon transition density matrix elements, $\hat{\mathcal{K}}_{n,\ve{x}=0;0,\ve{k},\ve{k}_0}$,that define the inhomogeneous source term are given by
 \beq \hat{\mathcal{K}}_{n,\ve{k};0,\ve{k}_0}=\la\la\Psi(n,\ve{x}=0 )\mid a_{\ve{k}}a^\dagger_{\ve{k}_0}\mid -\ve{k}_0,\psi_0 \ra\ra .\label{Kbarnkoko}\eeq
 In deriving these equations it has been assumed that the Hamiltonian $H$ has the structure, in a general single-nucleon basis $\mid i\ra$,
\beq H=\sum_{i,j}\la i\mid T\mid j \ra a_i^\dagger a_j+\frac{1}{4}\sum_{i,j.k.l}\la i,j\mid V_{\mathcal{A}}\mid k,l\ra a_i^\dagger a_j^\dagger a_l a_k,\label{H}\eeq
where $T$ is the single nucleon kinetic energy operator and the subscript $\mathcal{A}$ indicates that the matrix element of the nucleon-nucleon interaction $V$ involves normalised and antisymmetric two particle states $\mid i,j\ra$  and $\mid k,l\ra$ where $i,j,$ are labels of a complete set of single particle states. For example, for a state, $\mid i,j\ra_{(a,b)}$, of nucleons $a$ and $b$ an explicit form is 
\beq  \mid i,j\ra_{(a,b)}=\frac{1}{\sqrt{2}}(\mid i\ra_{(a)}\mid j\ra_{(b)}-\mid j\ra_{(a)}\mid i\ra_{(b)}). \label{ijstate}\eeq
The coupling potential $\la\ve{k},n  \mid\hat{ \mathcal{U}}\mid \ve{k}_1,n'  \ra$ that appears in eq.(\ref{scattwf4}) is defined in \cite{Johnson19}, eq.(59). It is related to the nucleon-nucleon interaction $V$ by
\beq \la\ve{k},n  \mid\hat{ \mathcal{U}}\mid \ve{k}_1,n'  \ra&&=\frac{1}{2}\int d\ve{k}_2\int d\ve{k}_4\la\ve{k},\ve{k}_2\mid V_{\mathcal{A}}\mid \ve{k}_1,\ve{k}_4\ra \la\la\Psi(n,\ve{x}=0)\mid a^\dagger_{\ve{k}_2}a_{\ve{k}_4}\mid-\ve{k}_1,\psi_{n'}\ra\ra \eol
&&=\frac{1}{2} \la\la\Psi(n,\ve{x}=0)\mid \mathcal{V}(\ve{k},\ve{k}_1)\mid-\ve{k}_1,\psi_{n'}\ra\ra ,\label{Uint}\eeq
where, in the notation of \cite{Johnson19}, eq.(46), the one-body operator $\mathcal{V}(\ve{k},\ve{k}_1)$ is 
\beq\mathcal{V}(\ve{k},\ve{k}_1)= \int d\ve{k}_2\int d\ve{k}_4\la\ve{k},\ve{k}_2\mid V_{\mathcal{A}}\mid \ve{k}_1,\ve{k}_4\ra  a^\dagger_{\ve{k}_2}a_{\ve{k}_4}.\label{Vkk12}\eeq
If a three-body force of the general form given in eq.(12) of \cite{Birse2020} is added to the Hamiltonian (\ref{H}), the operator $\mathcal{V}(\ve{k},\ve{k}_1)$ receives an additional contribution with the nature of an effective two-nucleon interaction and introduces two-nucleon target density matrix elements into the definition of $\hat{ \mathcal{U}}$ , see \cite{Birse2020}, eq.(13).
  
 It can be seen from eqs.(\ref{Psi1scatt2}) and (\ref{H}) that non-vanishing contributions from $V$ require the occupancy of states $\ve{k}_1$ and $\ve{k}_4$ in the $A+1$-nucleon  state $\mid \Psi_{E,\ve{k}_0}^\epsilon\ra\ra$. The summation over all choices for $\ve{k}_1$ and $\ve{k}_4$ includes both of the physically identical states $\mid \ve{k}_1,\ve{k}_4\ra$ and  $\mid \ve{k}_4,\ve{k}_1\ra$. This is corrected by the factor $\frac{1}{2}$ in the definition of $\hat{\mathcal{U}}$. The need for this factor is particularly transparent in the case $A=1$ discussed in Appendix \ref{A}.
\subsection{Introduction of $B$-space.}\label{A1B}
It is convenient here to recall the concept of $B$-space introduced following eq.(58) of \cite{Johnson19}. $B$-space consists of $A$-nucleons in  states  $\psi_n$ that interact with a fictitious particle of reduced mass $\frac{Am}{(A+1)}$  but  are not identical to it. A basis in this space is, by definition, an orthonormal set of states $\mid \ve{k},n \ra, n=0,\dots,\infty,\,\,\ve{k}$ arbitrary, where the fictitious particle has momentum $\ve{k}$ and the $A$ target nucleons are in an intrinsic state $n$ and have a total momentum $-\ve{k}$.  The interaction between the fictitious particle and the target is described by an operator $\hat{\mathcal{U}}$ in this extended space with matrix elements defined by eq.(\ref{Uint}). The "hat" notation over a quantity will be use to denote operators in $B$-space, as opposed to the bare-headed operators of Fock-space.

 The Hamiltonian of the system "fictitious particle+$A$ target nucleons" is
\beq\hat{\mathcal{H}}=\hat{T}+\hat{h}_A+\hat{\mathcal{U}},\label{HpartA}\eeq
where $\hat{T}$ is the kinetic energy operator associated with a particle of mass $\frac{A}{(A+1)}m$ and $\hat{h}_A$ is diagonal in the $\mid \ve{k},n \ra$ basis with eigenvalues $E_n$.
\beq \la \ve{k}',n'\mid\hat{h}_A \mid \ve{k},n \ra =\delta(\ve{k}-\ve{k}')\delta_{n,n'}E_n.\label{hA}\eeq

Using these concepts the coupled equations (\ref{scattwf4}) can be written 
\beq (E+\imath \epsilon-\hat{\mathcal{H}})\mid \hat{\Psi}_{\ve{k}_0,0}^{(\epsilon)}\ra=\imath \epsilon \hat{\mathcal{K}}(2\pi)^{3/2}\mid 0,\ve{k}_0\ra, \label{Beqcoupled}\eeq
where $\mid \hat{\Psi}_{\ve{k}_0,0}^{(\epsilon)}\ra$ is a ket in $B$-space with components
\beq \la \ve{k},n\mid \hat{\Psi}_{\ve{k}_0,0}^{(\epsilon)}\ra=\la\la\Psi(n,\ve{x}=0 )\mid a_{\ve{k}}\mid \Psi_{\ve{k}_0,0}^{(\epsilon)}\ra\ra.\label{Psihat}\eeq
\subsection{Interpretation of the coupled equations. The optical model wave function.}\label{interp}
In \cite{Johnson17}, the optical model wave function, $ \xi_{E,\ve{k}_0}^\epsilon(\ve{r})$, corresponding to the elastic scattering of a nucleon of momentum $\ve{k}_0$ in the overall c.m. system by an $A$-nucleon target in its ground state was formally defined as a matrix element between many-nucleon states  in Fock-space through the formula
 \beq \xi_{E,\ve{k}_0}^\epsilon(\ve{r})=\la\la \Psi(0, \ve{x}=0) \mid \psi(\ve{r})\mid \Psi_{E,\ve{k}_0}^\epsilon\ra\ra. \label{xiEepsdef}\eeq
The operator $\psi(\ve{r})$ destroys a nucleon at a point labelled $\ve{r}$. The equivalent expression in momentum space is
 \beq \xi_{E,\ve{k}_0}^\epsilon(\ve{k})=\la\la \Psi(0, \ve{x}=0) \mid a_{\ve{k}}\mid \Psi_{E,\ve{k}_0}^\epsilon\ra\ra. \label{xiEepsdef2}\eeq
 Comparing with the definition  (\ref{Psihat}) it is clear that the optical model wavefunction is the  $n=0$ component of the ket in $B$-space  that is the solution of the coupled equations (\ref{Beqcoupled}). These equations are reminiscent of the coupled equations used in Feshbach's theory  of the optical model \cite{Fes58} but with two important differences.
 
 (1) The full effects of antisymmetrisation and recoil are included in the coupling potential $\hat{\mathcal{U}}$ defined in eq.(\ref{Uint}) through direct and exchange matrix elements of the nucleon-nucleon interaction $V$ and the the occurrence in eq.(\ref{Uint}) of one-nucleon density matrix elements $\la\la\Psi(n,\ve{x}=0)\mid a^\dagger_{\ve{k}_2}a_{\ve{k}_4}\mid-\ve{k}_1,\psi_{n'}\ra\ra$  in a mixed basis with a localised target centre-of-mass in the bra and a definite target centre-of-mass momentum in the ket. 
 
 (2) The non-standard form of the source term $\imath \epsilon \hat{\mathcal{K}}(2\pi)^{3/2}\mid 0,\ve{k}_0\ra$ that appears on the right in eq.(\ref{Beqcoupled}). Standard scattering theory would have a source term $\imath \epsilon(2\pi)^{3/2}\mid 0,\ve{k}_0\ra $, i.e.,  $\hat{\mathcal{K}}=1 $, and in the limit $\epsilon \rightarrow 0^+$ produce a state  with a plane wave of unit amplitude in the incident channel plus outgoing waves in all channels. It is shown in Appendix \ref{B} that  terms in the operator $\hat{\mathcal{K}}$ that deviate from unity give zero contribution to $\mid \Psi_{\ve{k}_0,0}^{(\epsilon)}\ra $ in the limit $\epsilon \rightarrow 0^+$ and the coupled equations (\ref{Beqcoupled}) can be replaced by
 \beq (E+\imath \epsilon-\hat{\mathcal{H}})\mid \hat{\Psi}_{\ve{k}_0,0}^{(\epsilon)}\ra=\imath \epsilon (2\pi)^{3/2}\mid 0,\ve{k}_0\ra. \label{Beqcoupled2}\eeq
In the 3-body case, i.e., $A=2$ it may be necessary to take the limit $\epsilon \rightarrow 0^+$ by converting these equations to the set of coupled Faddeev \cite{Fad61} or AGS \cite{AGS67} equations. 
  \subsection{Feshbach's theory of the optical potential applied to $B$-space overlap functions.}\label{FeshbachTheory}
In this Section the Feshbach \cite{Fes58} theory of the optical potential is reviewed in a form that is convenient for later discussions of the connection between nucleon optical potential theory and the effective interaction used in 3-body models of $A(d,p)B$ reactions.

In the notation of eq.(\ref{HpartA})
\beq\hat{\mathcal{H}}=\hat{\mathcal{H}}_0+\hat{\mathcal{U}},\label{HpartA2}\eeq
where
\beq\hat{\mathcal{H}}_0=\hat{T}+\hat{h}_A.\label{HpartA3}\eeq 

The objective of the following algebra is to derive uncoupled equations for the components $ P_0\mid \hat{\Psi}_{\ve{k}_0,0}^{(\epsilon)}\ra $ and \newline $ Q_0\mid \hat{\Psi}_{\ve{k}_0,0}^{(\epsilon)}\ra $ where $P_0$ projects onto states $\mid \ve{k}, 0\ra$ in $B$-space with the target is in its ground state, $n=0$, and $Q_0$ projects onto the orthogonal sub-space of states with $n\neq 0.$ 
Using these projection operators, eq.(\ref{Beqcoupled}) can be written
\beq (E+\imath \epsilon-\hat{\mathcal{H}_0}-P_0\hat{\mathcal{U}}P_0)P_0\mid \hat{\Psi}_{\ve{k}_0,0}^{(\epsilon)}\ra=&&\imath \epsilon (2\pi)^{3/2}\mid 0,\ve{k}_0\ra+P_0\hat{\mathcal{U}}Q_0\mid \hat{\Psi}_{\ve{k}_0,0}^{(\epsilon)}\ra,\eol
(E+\imath \epsilon-\hat{\mathcal{H}_0})Q_0\mid \hat{\Psi}_{\ve{k}_0,0}^{(\epsilon)}\ra=&&Q_0\hat{\mathcal{U}}Q_0\mid \hat{\Psi}_{\ve{k}_0,0}^{(\epsilon)}\ra+Q_0\hat{\mathcal{U}}P_0\mid \hat{\Psi}_{\ve{k}_0,0}^{(\epsilon)}\ra.\label{Beqcoupled3}\eeq
Solving the second of eqs.(\ref{Beqcoupled3}) for $Q_0\mid \hat{\Psi}_{\ve{k}_0,0}^{(\epsilon)}\ra$ and substituting the result into the first equation gives the following equation for $P_0\mid \hat{\Psi}_{\ve{k}_0,0}^{(\epsilon)}\ra$: 
 \beq (E+\imath \epsilon-\hat{\mathcal{H}_0}-P_0\hat{\mathcal{U}}P_0-P_0\hat{\mathcal{U}}Q_0\frac{1}{(E+\imath \epsilon-\hat{\mathcal{H}_0}-Q_0\hat{\mathcal{U}}Q_0)})Q_0\hat{\mathcal{U}}P_0\mid \hat{\Psi}_{\ve{k}_0,0}^{(\epsilon)}\ra=&&\imath \epsilon (2\pi)^{3/2}\mid 0,\ve{k}_0\ra.\eol
.\label{Beqcoupled4}\eeq
This result identifies the optical potential as the ground state matrix element of the effective interaction operator in $B$-space, $\hat{U}_{\mathrm{eff}}$, defined by
 \beq \hat{U}_{\mathrm{eff}}=\hat{ \mathcal{U}}+\hat{\mathcal{U}}Q_0\frac{1}{(E+\imath \epsilon-\hat{T}-\hat{h}_A-Q_0\hat{ \mathcal{U}}Q_0)}Q_0\hat{ \mathcal{U}}, \label{Ueff}\eeq
 which is the formal solution to the equation
  \beq \hat{U}_{\mathrm{eff}}=\hat{ \mathcal{U}}+\hat{\mathcal{U}}Q_0\frac{1}{(E+\imath \epsilon-\hat{T}-\hat{h}_A)}Q_0\hat{U}_{\mathrm{eff}}. \label{Ueff2}\eeq
 The optical potential operator in momentum space is
 \beq \la\ve{k}\mid\hat{V}_{\mathrm{opt}}\mid \ve{k}'\ra=\la\ve{k},0\mid\hat{U}_{\mathrm{eff}}\mid \ve{k}',0\ra.\label{Vhatopt}\eeq
 Note that in the expression on the right the matrix elements of $\hat{ \mathcal{U}}$ that appear in the evaluation of $\hat{U}_{\mathrm{eff}}$ must be interpreted using the relations defined in eqs.(\ref{Uint}) and (\ref{Vkk12}) in order to preserve translation invariance.

 \section{Deuteron elastic and inelastic scattering and break-up on an $A$-nucleon target.}\label{dAscattering}
 The following Sections explore  a fully antisymmetrised, translationally invariant theory of the $n+p+A$ system in a form that enables a link to be made with the theory of the nucleon optical model described above and hence with 3-body models of deuteron stripping reactions.                                                                                                                                           
\subsection{Deuteron creation and  destruction operators.}\label{deutdagger}
Following Villars \cite{Villars1967}, the operator that creates a deuteron with total momentum $\ve{K}_{d_0}$ and intrinsic angular momentum projection  $M_{d_0}=\pm 1,0$ is 
\beq A^\dagger_{d_0,\ve{K}_{d_0},M_{d_0}}=\frac{1}{\sqrt{2}}\int d\,1 d\,2 \la 1,2 \mid d_0,\ve{K}_{d_0},M_{d_0}\ra a^\dagger_1a^\dagger_2,
\label{Adaggerd}\eeq
where $1,2,\dots ,$ denote  possible values of a set of space, intrinsic spin ($\sigma $) and isospin($\tau$) coordinates  and the $a_i,a^\dagger_j,$ are nucleon operators.. The function $\la 1,2\mid d_0, \ve{K}_{d_0},M_{d_0}\ra$ describes the deuteron ground state with total momentum $\ve{K}_{d_0}$. In a basis in which the space degree of freedom is chosen to be nucleon position, it has the form  
\beq \la 1,2\mid d_0, \ve{K}_{d_0},M_{d_0}\ra=\frac{1}{(2\pi)^{3/2}}\exp(\imath \ve{K}_{d_0}.((\ve{r}_1+\ve{r}_2)/2)\phi_{d_0,M_{d_0}}((\ve{r}_1-\ve{r}_2),\sigma_1,\sigma_2)\chi_{0,0}(\tau_1,\tau_2).\label{KS}\eeq
    
   A destruction operator corresponding to eq.(\ref{Adaggerd}) is given by
\beq A_{d_0,\ve{K}'_{d_0},M'_{d_0}}=\frac{1}{\sqrt{2}}\int d\,3 d\,4( \la 3,4 \mid d_0, \ve{K}'_{d_0},M'_{d_0}\ra)^* a_4a_3,
\label{Ad}\eeq
where the ordering of the labels $3,\,4$ should be noted.

The states $A^\dagger_{d_0,\ve{K}_{d_0},M_{d_0}}\mid 0\ra\ra $ are normalised so that
 \beq \la\la 0 \mid A_{d_0,\ve{K'_{d_0}},M'_{d_0}}A^\dagger_{d_0,\ve{K}_{d_0},M_{d_0}}\mid 0 \ra\ra=\delta(\ve{K}'_{d_0}-\ve{K}_{d_0})\delta_{M_{d_0},M'_{d_0}},
\label{AKorth11}\eeq
where $\mid 0 \ra\ra$ is the nucleon vacuum.

Note that 
\beq [ A_{d_0,\ve{K'_{d_0}},M'_{d_0}},A_{d_0,\ve{K}_{d_0},M_{d_0}}]_-= [A^\dagger_{d_0,\ve{K'_{d_0}},M'_{d_0}},A^\dagger_{d_0,\ve{K}_{d_0},M_{d_0}}]_-=0, \label{Bosoncomms}\eeq
as expected if boson creation and destruction operators were involved. These follow trivially from $[a_4a_3,a_1a_2]_=0$, etc. However the other boson commutation relation is not generally valid, i.e.,
\beq [ A_{d_0,\ve{K'_{d_0}},M'_{d_0}},A^\dagger_{d_0,\ve{K}_{d_0},M_{d_0}}]_-\neq \delta(\ve{K}'_{d_0}-\ve{K}_{d_0})\delta_{M'_{d_0},M_{d_0}.} \label{Bosoncomms2}\eeq
The equality sign is only valid when the the pair of two-nucleon systems are in a state where they are spatially separated by distances greater than the range of nucleon-nucleon forces or when acting on the vacuum. The correct value of the anticommutator is given in the Appendix C.  
\subsection{Generalisation to include  $n-p$ break-up states.}\label{General n-p}
The above analysis is easily generalised to include continuum states of the two-nucleon system.  A complete set of antisymmetrised states  $\mid d (a,b) \ra$ of two nucleons labelled $a$ and $b$ and definite total momentum $\ve{K}_d $ is introduced. The label $d$ ranges over all possible values, continuum and discrete, of a complete set of eigenvalues needed to specify eigenstates of the two-nucleon Hamiltonian. Among these states will be the deuteron ground state and a set of nucleon-nucleon scattering states. The labels $d,d', \dots, $ will be used to designate any of these state, with $d_0$ used if the deuteron ground state is meant.  The states $\mid d (a,b) \ra$ will be assumed to be orthonormalised and complete, i.e.,
\beq \la d(a,b)\mid d'(a,b)\ra=\delta_{d,d'},\,\,\,\, \sum_d \mid d(a,b) \ra\la d(a,b)\mid =\mathcal{I},\label{phisprops}\eeq 
where $\mathcal{I}$ is the unit operator in the space of antisymmetrised two-nucleon states.

The generalisation of eq.(\ref{KS}) for the continuum states is
\beq \la \ve{r}_1,\sigma_1,\tau_1;\ve{r}_2,\sigma_2,\tau_2\mid d, \ve{K}_{d},\ve{k}_d,M_{d},T_d,\tau_d\ra=\frac{\exp(\imath \ve{K}_{d}.((\ve{r}_1+\ve{r}_2)/2)}{(2\pi)^{3}}\phi^{(+)}_{\ve{k}_d,M_{d},T_d}((\ve{r}_1-\ve{r}_2),\sigma_1,\sigma_2)\chi_{T_d,\tau_d}(\tau_1,\tau_2),\label{KS2}\eeq
where $\sigma_i,\tau_i$ are, respectively,  nucleon intrinsic spin and isospin coordinates. The function $\phi^{(+)}_{\ve{k}_d,M_{d},T_d}$ is a nucleon-nucleon scattering state in the nucleon-nucleon centre-of-mass system corresponding to an incident wave of  momentum $\ve{k}_d$ and unit amplitude, isospin $T_d=0,1$ and where $M_d$ is a set of intrinsic spin quantum numbers specifying the incident channel.

The corresponding state in a single-nucleon momentum basis can be written
\beq \la \ve{k}_1,\sigma_1,\tau_1;\ve{k}_2,\sigma_2,\tau_2\mid d, \ve{K}_{d},\ve{k}_d,M_{d},T_d,\tau_d\ra=\delta(\ve{k}_1+\ve{k}_2-\ve{K}_d)\phi^{(+)}_{\ve{k}_d,M_{d},T_d}((\ve{k}_1-\ve{k}_2)/2,\sigma_1,\sigma_2)\chi_{T_d,\tau_d}(\tau_1,\tau_2).\label{KS3}\eeq
  
A general creation operator will be written
\beq A^\dagger_{d}=\frac{1}{\sqrt{2}}\int d\,1 d\,2 \la 1(a),2(b) \mid d(a,b)\ra a^\dagger_1a^\dagger_2.
\label{Adaggerdgen4}\eeq
Here $\mid 1(a),2(b) \ra$ denotes one of the  two-nucleon states
\beq \mid 1(a),2(b)\ra=\mid 1(a)\ra \mid 2(b)\ra \label{12} \eeq
where the ket $\mid 1(a)\ra$ is a one nucleon state of nucleon $a$  with the label $1$ being one of a collection of space, spin and isospin labels needed to define a one-nucleon state.   The states  $\mid 1(a),2(b) \ra$ satisfy
\beq \la 1(a),2(b)\mid 1'(a),2'(b)\ra =\delta_{1,1'} \delta_{2,2'}, \label{12norm}\eeq
and

\beq \int\,d1\mid 1(a)\ra\la 1(a) \mid=1.\label{12complete}\eeq

These states should be distinguished from the antisymmetrised two-nucleon states, $\mid 1,2 \ra$, that appear, for example, in the definition of the two-nucleon matrix elements in eq.(\ref{ijstate}). The non-symmetrised states (\ref{12}) will prove to be useful later in the discussion of 3-body models of the $d+A$ system.  

 The function $\la 1(a),2(b)\mid d(a,b)\ra$ satisfies
\beq \la 1(a),2(b)\mid (E_{d}-T_1-T_2)\mid d(a,b)\ra=\int d3\,d4\,\frac{1}{2}\la 1,2\mid V_{\cal{A}}\mid 3,4\ra\la 3(a),4(b)\mid d(a,b)\ra, \label{phdeq}\eeq
where the energy $E_{d}$ is given by 
\beq E_{d_0}=\frac{\hbar^2K_{d_0}^2}{4m}-\epsilon_{d_0}, \label{Ed0def}\eeq
where $m$ is the nucleon mass and $\epsilon_{d_0}$ is the deuteron binding energy and for a general continuum state
\beq E_{d}&&=\frac{\hbar^2K_d^2}{4m}+\frac{\hbar^2k_d^2}{m}\eol
&&=\frac{\hbar^2K_d^2}{4m}+\epsilon_d. \label{Eddef}\eeq
In the following  $\epsilon_d$ will mean $\frac{\hbar^2k_d^2}{m}$ when $d\neq d_0$ and $\epsilon_d=-\epsilon_{0}$ when $d=d_0$.

The assumed properties of the kets $\mid \phi_d \ra$ given in eqs.(\ref{phisprops}) together with eqs.(\ref{12norm}) and (\ref{12complete}) give
\beq \int d1d2 \la d(a,b) \mid 1(a),2(b)\ra \la 1(a),2(b)\mid d'(a,b)\ra=\delta_{d,d'}.\label{12phid2}\eeq
The inverse of eq.(\ref{Adaggerdgen4}) that follows from (\ref{12phid2})  is
 \beq  a^\dagger_1a^\dagger_2=\sqrt{2}\sum_d\la d(a,b)\mid 1(a),2(b)\ra A^\dagger_{d}.
\label{Adaggerdgen3}\eeq

\subsection{Coupled equations for $d+A$ overlap amplitudes.}\label{Jd}
 The aim here is to derive coupled equations for the overlaps that describe $d+A$ scattering that are the analogues of equations (\ref{scattwf4}) for nucleon$+A$ scattering.  
 
  In the overall c.m. system, the antisymmetrized $A+2$-nucleon scattering wave function corresponding to a deuteron incident on an $A$-nucleon target in its ground state is the limit $\epsilon \rightarrow 0^+$ of the state $\mid \Psi^{\epsilon}_{d_0,0}\ra\ra$  defined by
\beq (E+\imath \epsilon-H)\mid \Psi^{\epsilon}_{d_0,0}\ra\ra&&=\imath\epsilon (2\pi)^{3/2} A^\dagger_{d_0}\mid -\ve{K_{d_0}}, \psi_{0}\ra\ra.  
\label{dAscatt}\eeq
Proceeding as in the nucleon$+A$ case, inner products are taken with both sides of this equation with the kets \newline $A^\dagger_{d''}\mid -\ve{K}_{d}, \psi_{n}\ra\ra$ and using an appropriate expression for the commutator $[A_{d},H]_-$. It is  shown in Appendix \ref{D} that  
\beq [A_d,H]_-=E_{d}A_d+\sum_{d'}\mathcal{V}_{d,d'} A_{d'},\label{Jdef}\eeq
where $E_d$ is defined in eq.(\ref{Eddef}) and
\beq \mathcal{V}_{d,d'}= \int \,d2\,d3\,\rho_{d,d'}(3,2)\mathcal{V}(3, 2), \label{Vd'd32} \eeq
 where $\mathcal{V}(3, 2)$ is defined in eq.(\ref{Vkk12}) and
  \beq \rho_{d,d'}(3,2)=\int\,d1\,\la d(a,b)\mid 1(a),3(b)\ra \la 1(a),2(b) \mid d'(a,b)\ra,\label{rhodd'}\eeq
 is a matrix of one-nucleon transition density matrices associated with the two-nucleon states $d$ and $d'$.
 
Hence
\beq \la\la-\ve{K}_{d}, \psi_{n}\mid A_{d"}(E+\imath \epsilon-H)\mid \Psi^{\epsilon}_{d_0,0}\ra\ra&&=\imath\epsilon (2\pi)^{3/2} \la\la-\ve{K}_{d}, \psi_{n}\mid A_{d"} A^\dagger_{d_0}\mid -\ve{K_{d_0}}. \psi_{0}\ra\ra.  
\label{dAscatt2}\eeq
Using eq.(\ref{Jdef}) and the result
\beq H\mid -\ve{K}_{d}, \psi_{n}\ra\ra=(E_n+\frac{\hbar^2K_{d}^2}{2Am})\mid -\ve{K}_{d}, \psi_{n}\ra\ra,\label{HKdn}\eeq
eq.(\ref{dAscatt2}) becomes
 \beq (E+\imath \epsilon-E_n
 -\frac{\hbar^2K_{d}^2}{2Am}-E_d)\la\la-\ve{K}_{d}, \psi_{n}\mid A_{d"}\mid \Psi^{\epsilon}_{d_0,0}\ra\ra-\!\!\!\!\!&&\sum_{d'''}\la\la-\ve{K}_{d}, \psi_{n}\mid \mathcal{V}_{d'',d'''} A_{d'''}\mid \Psi^{\epsilon}_{d_0,0}\ra\ra\!\!\!\!\!\eol&&=\imath\epsilon (2\pi)^{3/2} \la\la-\ve{K}_{d}, \psi_{n}\mid A_{d''} A^\dagger_{d_0}\mid -\ve{K_{d_0}}. \psi_{0}\ra\ra,\eol&&  
\label{dAscatt3}\eeq
where $E_d$ is given by eq.(\ref{Eddef}). Introducing a complete set of $A$-nucleon states $\mid -\ve{K}_{d'},\psi_{n'}\ra\ra $,  the second term on the left in (\ref{dAscatt3}) can be written 
\beq \sum_{d'''}\la\la-\ve{K}_{d}, \psi_{n}\mid \mathcal{V}_{d'',d'''} A_{d'''}\mid \Psi^{\epsilon}_{d_0,0}\ra\ra=\frac{1}{(2\pi)^3}\sum_{d',n'}\sum_{d'''}\la\la-\ve{K}_{d}, \psi_{n}\mid \mathcal{V}_{d",d'''}\mid -\ve{K}_{d'},\psi_{n'}\ra\ra\la\la  -\ve{K}_{d'},\psi_{n'}\mid A_{d'''}\mid \Psi^{\epsilon}_{d_0,0}\ra\ra .  
\eol&&\label{dAscatt4}\eeq
These equations can be considerably simplified by using momentum conservation for a translationally invariant nucleon-nucleon interaction $V$ and the properties of various state of definite momentum that appear in eqs.(\ref{dAscatt3}) and (\ref{dAscatt4}).

In the first place, the definition of $A^\dagger_d$ means that increases the total momentum of any state it acts on by $\ve{K}_d$. The overlap $\la\la-\ve{K}_{d'}, \psi_{n'}\mid A_{d'''}\mid \Psi^{\epsilon}_{d_0,0}\ra\ra$ must be proportional to $\delta(\ve{K}_{d'}-\ve{K}_{d'''})$ because $\mid \Psi^{\epsilon}_{d_0,0}\ra\ra$ has total momentum zero. It follows that
\beq \la\la-\ve{K}_{d'}, \psi_{n'}\mid A_{d'''}\mid \Psi^{\epsilon}_{d_0,0}\ra\ra=(2\pi)^3\delta(\ve{K}_{d'}-\ve{K}_{d'''})\la\la \Psi(n',\ve{x}=0\mid A_{d'}\mid \Psi^{\epsilon}_{d_0,0}\ra\ra, \label{overlap}\eeq
and eq.(\ref{dAscatt4})  reduces to
\beq \sum_{d'''}\la\la-\ve{K}_{d}, \psi_{n}\mid \mathcal{V}_{d'',d'''} A_{d'''}\mid \Psi^{\epsilon}_{d_0,0}\ra\ra=\sum_{n',d'}\la\la-\ve{K}_{d}, \psi_{n}\mid \mathcal{V}_{d'',d'}\mid -\ve{K}_{d'},\psi_{n'}\ra\ra\la\la \Psi(n',\ve{x}=0)\mid A_{d'}\mid \Psi^{\epsilon}_{d_0,0}\ra\ra .  
\eol&&\label{dAscatt5}\eeq
Secondly, for a translational invariant $V$, a similar treatment gives
 \beq  \la\la -\ve{K}_d,\psi_n\mid\mathcal{V}_{d'',d'}\mid -\ve{K}_{d'},\psi_{n'} \ra\ra=(2\pi)^3\delta(\ve{K}_{d}-\ve{K}_{d''})\bar{\mathcal{U}}_{d,n;d',n'}, \eol &&\label{Vd'd32ME2} \eeq
 where 
 \beq \bar{\mathcal{U}}_{d,n;d',n'}&&=(2\pi)^3\int\,d1\,d2\,d3\,\delta(\ve{k}_1+\ve{k}_{3}-\ve{K}_{d})\delta(\ve{k}_1+\ve{k}_2-\ve{K}_{d'})\eol
\eol&& \times \la\Psi(d,\ve{x}=0)\mid 1(a),3(b)\ra \la 1(a),2(b) \mid \Psi(d',\ve{x}=0)\ra \la\la  \Psi(n,\ve{x}=0)\mid \mathcal{V}(\ve{k}_3, \ve{k}_2)
 \mid -\ve{K}_{d'}, \psi_{n'}\ra\ra.\label{Ubardnd'n'}\eeq
  In deriving these results it has been assumed that the space part of the nucleon single-particle state labels $1,2,3,\dots,$ have been chosen to be the momenta $\ve{k}_1,\ve{k}_2,\ve{k}_3,\dots$. The integration over the momentum  $\ve{k}_1$ implied by $\int\,d1$, in eq.(\ref{Ubardnd'n'}) has been retained, although of course this integration could be eliminated using the delta functions. This makes  clear that there is still a summation over the spin and isospin labels associated with state $1$.
  
  The interaction $\bar{\mathcal{U}}$ places a role here that is analogous to the interaction $\hat{\mathcal{U}}$ in the $A$+1 case. It couples states in the two nucleon sub-space spanned by the states $d,d',\dots$ and $A$-nucleon target states $n,n',\dots$.
         
 Assembling the results (\ref{overlap})-(\ref{Vd'd32ME2}), eq.(\ref{dAscatt3}) reduces to a set of coupled equations for the amplitudes $\la\la \Psi(n,\ve{x}=0)\mid A_{d}\mid \Psi^{\epsilon}_{d_0,0}\ra\ra$ of the form
 \beq (E+\imath \epsilon-E_n
 -\frac{\hbar^2K_{d}^2}{2Am}-E_d)\la\la \Psi(n,\ve{x}=0)\mid A_{d}\mid \Psi^{\epsilon}_{d_0,0}\ra\ra-\!\!\!\!\!&&\sum_{d',n'}\bar{\mathcal{U}}_{d,n;d',n'}\la\la \Psi(n',\ve{x}=0)\mid A_{d'}\mid \Psi^{\epsilon}_{d_0,0}\ra\ra\!\!\!\!\!\eol&&=\imath\epsilon (2\pi)^{3/2} \la\la\Psi(n,\ve{x}=0)\mid A_{d} A^\dagger_{d_0}\mid -\ve{K_{d_0}}, \psi_{0}\ra\ra,\eol&&  
\label{dAscatt5}\eeq
Using the result proved in Appendix \ref{C} for the commutator $[ A_{d'},A_d^\dagger\,] $ the source term on the right of eq.(\ref{dAscatt5}) can be written
\beq  \la\la\Psi(n,\ve{x}=0)\mid A_{d} A^\dagger_{d_0}\mid -\ve{K_{d_0}}, \psi_{0}\ra\ra&&=\delta_{d,d_0}\delta_{n,0}\eol
&&-2\int d\,1 d\,2\, d\,4   \la \phi_{d}\mid 1(a),4(b)\ra \la 1(a),2(b) \mid\phi_{d_0}\ra \la\la\Psi(n,\ve{x}=0)\mid a^\dagger_2a_4\mid -\ve{K_{d_0}}, \psi_{0}\ra\ra.\eol&&
 \label{inhomd}\eeq
 Using arguments similar to those discussed in Section \ref{interp} for the nucleon$+A$ case it is shown in Appendix \ref{F} that if $\phi_{d_0}$ and $\psi_0$ are stable ground states the non-delta function terms on the right in eq.(\ref{inhomd}) give a vanishing contribution to the solution of Eq.(\ref{dAscatt5}) in the limit $\epsilon \rightarrow 0 $. Eq.(\ref{dAscatt5}) can therefore be replaced by  
 
 \beq (E+\imath \epsilon-E_n
 -\frac{\hbar^2K_{d}^2}{2Am}-E_d)\la\la \Psi(n,\ve{x}=0\mid A_{d}\mid \Psi^{\epsilon}_{d_0,0}\ra\ra-\!\!\!\!\!&&\sum_{d',n'}\bar{\mathcal{U}}_{d,n;d',n'}\la\la \Psi(n',\ve{x}=0\mid A_{d'}\mid \Psi^{\epsilon}_{d_0,0}\ra\ra\!\!\!\!\!\eol&&=\imath\epsilon (2\pi)^{3/2} \delta_{d,d_0}\delta_{n,0}.\eol&&  
\label{dAscatt6}\eeq
\section{Three-body model Hamiltonian in $\bar{B}$-space for the $n+p+A$ system. }\label{BdA}
   It is convenient to introduce the concept of $\bar{B}$-space for the $n+p+A$ system by analogy with the $B$-space formalism used in  for  nucleon$+A$ scattering in Section \ref{A1B} above.
   
    An  $A-$nucleon target with  states labelled  $n,n',\dots$ with intrinsic energies $E_n,E_{n'},\dots ,$ interacts with a fictitious particle of  mass $\frac{2Am}{(A+2)}$ and internal states $d $ with internal energies $\epsilon_d$  in the convention introduced at the beginning of Section \ref{General n-p}. 
    
     A basis in this space is, by definition, an orthonormal set of states $\mid \ve{K}_d,\epsilon_d, n \ra,$  where the fictitious particle has momentum $\ve{K}_d$ and the target nucleons have an intrinsic state $n$ and a total momentum $-\ve{K}_d$. The state $\mid \ve{K}_d,\epsilon_d, n \ra$ will frequently be denoted simply  $\mid d,n\ra$. 
     
     The constituents of the fictitious particle in the $d$-states are treated as not identical to the target nucleons. The fictitious particle interacts with $A$  through the  operator $\bar{\mathcal{U}}$ in the $\bar{B}$ space with matrix elements defined in terms of the Fock-space elements of $\mathcal{V}$  by eq.(\ref{Ubardnd'n'}). The interaction $\mathcal{V}$ includes knock-on exchange effects arising from the identity of the nucleons in the target and nucleons in the fictitious particle. These relations are reproduced here for convenience.
    \beq \bar{\mathcal{U}}_{d,n;d',n'}&&=(2\pi)^3\int\,d1\,d2\,d3\,\delta(\ve{k}_1+\ve{k}_{3}-\ve{K}_{d})\delta(\ve{k}_1+\ve{k}_2-\ve{K}_{d'})\eol
\eol&& \times \la\Psi(d,\ve{x}=0)\mid 1(a),3(b)\ra \la 1(a),2(b) \mid \Psi(d',\ve{x}=0)\ra \la\la  \Psi(n,\ve{x}=0)\mid \mathcal{V}(\ve{k}_3, \ve{k}_2)
 \mid -\ve{K}_{d'}, \psi_{n'}\ra\ra.\label{Ubardnd'n'2}\eeq 
 The Hamiltonian of the three-body system "fictitious particle+$A$ target nucleons" is
\beq\bar{\mathcal{H}}&&=\bar{T}+\bar{h}_{A}+\bar{h}_2+\bar{\mathcal{U}}\eol
&&=\bar{\mathcal{H}}_0+\bar{\mathcal{U}},\label{HbartA}\eeq
where
\beq \bar{\mathcal{H}}_0= \bar{T}+\bar{h}_{A}+\bar{h}_2,\label{Hbar0}\eeq
and $\bar{T}$ is the kinetic energy operator associated with a particle of mass $\frac{2A}{(A+2)}m$ and $\bar{h}_{A}$ and $\bar{h}_2$ are diagonal in the $\mid d,n \ra$ basis with eigenvalues $E_n$ and $\epsilon_d$, respectively.
\beq \la d',n'\mid(\bar{h}_{A}+\bar{h}_2) \mid d,n \ra =\delta_{d',d}\delta_{n',n}(E_n+\epsilon_d).\label{h(A+2)}\eeq
The coupled equations (\ref{dAscatt6}) can be written as an equation in $\bar{B}$-space. 
 \beq (E+\imath \epsilon -\bar {\mathcal{H}})\mid \Psi^{(\epsilon)}_{d_0,0}\ra=\imath \epsilon \mid d_0,0\ra.\label{Beq}\eeq
 The solution of this equation is a column vector, $\mid \Psi^{(\epsilon)}_{d_0,0}\ra $ in $\bar{B}$-space with rows labelled by all possible values of the set $d,n$. The entry in the row labelled $d,n$ is the overlap $\la\la \Psi(n,\ve{x}=0\mid A_{d}\mid \Psi^{\epsilon}_{d_0,0}\ra\ra$. Operators in $\bar{B}$-space are shown with a bar above a symbol to distinguish them from operators in $B$-space which have hats.

\subsection{Structure of the three-body Hamiltonian. }\label{3bodyHamiltonian}
The interaction term $\bar{\mathcal{U}}$ in the model Hamiltonian $\bar {\mathcal{H}}$ is given by (eq.(\ref{Ubardnd'n'})
 \beq \bar{ \mathcal{U}}_{d',n':d,n}= (2\pi)^3\int\,d1\,d2\,d3\,\delta(\ve{k}_1+\ve{k}_{3}-\ve{K}_{d})\!\!\!\!\!\!\!&&\delta(\ve{k}_1+\ve{k}_2-\ve{K}_{d'})\la\Psi(d',\ve{x}=0)\mid 1(a),3(b)\ra \la  1(a),2(b) \mid \Psi(d,\ve{x}=0)\ra \eol \times\!\!\!\!\!\!\!\! &&\la\la  \Psi(n',\ve{x}=0)\mid \mathcal{V}(3, 2)
 \mid -\ve{K}_d, \psi_n \ra\ra,\label{Ubardnd'n'2}\eeq
 It is clear from this expression that $\bar{\mathcal{U}}$ has the form of a folding of a sum of operators where each operator involves just one of the two nucleons in the fictitious particle (nucleon $b$) interacting with the target nucleons through a two-nucleon interaction. The latter interaction includes exchange terms in $\mathcal{V}(3, 2)$. The way these terms appear in (\ref{Ubardnd'n'2}) is similar to the operator $\hat{\mathcal{U}}$  defined in eq.(\ref{Uint}) in the theory of the nucleon optical potential described in Section \ref{nucleonopt} and reproduced here for convenience:
 \beq \la\ve{k},n  \mid\hat{ \mathcal{U}}\mid \ve{k}',n'  \ra=\frac{1}{2}\la\la\Psi(n,\ve{x}=0) \mid\mathcal{V}(\ve{k}, \ve{k}') \mid-\ve{k}',\psi_{n'}\ra\ra,\label{Uint2}\eeq
  The kinematics involved in the $(A+2)$-nucleon and the $A$-nucleon problems differ in detail and make the relation between the interactions that appear in the three-body model and  nucleon optical potentials more complicated (see Appendix 1 of \cite{Tim13} for an example of this). The matrix element of $ \mathcal{V}(3, 2)$ that appears in eq.(\ref{Ubardnd'n'2}) involves three independent momenta whereas in eq.(\ref{Uint2}) only two are involved. However, both $\bar{\mathcal{U}}$ and $\hat{\mathcal{U}}$ are  related to the matrix elements
  \beq \mathcal{U}_{(n,0;n',\ve{x})}(\ve{k},\ve{k}')=\frac{1}{2}\la\la\Psi(n,\ve{x}=0) \mid\mathcal{V}(\ve{k}, \ve{k}') \mid\Psi(n',\ve{x})\ra\ra.\label{xrepME}\eeq 

\subsection{Physical interpretation of the interaction in the three-body Hamiltonian.}
  
To elucidate the interaction (\ref{Ubardnd'n'2})  further it is convenient to have an expression for  $\bar{\mathcal{U}}$ in a $\bar{B}$-space basis that differs from the $d$ basis and instead displays the individual degrees of freedom of the two-nucleons forming the fictitious particle are displayed. This can be achieved by evaluating  
 (note that, e.g., $\sum_d$ includes an integration over $\ve{K}_d$)
 \beq \bar{\mathcal{U}}_{n,n'}=\sum_{d',d}\frac{1}{(2\pi)^3}\mid \ve{K}_{d'},d'\ra \bar{ \mathcal{U}}_{d',n':d,n} \la \ve{K}_{d},d \mid. \label{UbarSP}\eeq
For a fixed $n,n'$ this defines an operator in the antisymmetrised two-nucleon  fictitious particle subspace of $\bar{B}$-space. 

To carry out the $d,d'$ summation in eq,(\ref{UbarSP}) it is convenient  to write
\beq \int\,d1&&\delta(\ve{k}_1+\ve{k}_{3}-\ve{K}_{d})\delta(\ve{k}_1+\ve{k}_2-\ve{K}_{d'})\la\Psi(d',\ve{x}=0)\mid 1(a),3(b)\ra \la 1(a),2(b) \mid \Psi(d,\ve{x}=0)\ra\eol &&= \int\,d1\frac{1}{(2\pi)^6}\la  \ve{K}_{d'},d'\mid 1(a),3(b)\ra\la 1(a),2(b)  \mid  \ve{K}_{d},d\ra. \label{rhod'd}\eeq
Using this and the completeness of the states $\mid   \ve{K}_{d'},d'\ra$ the sum over $d'$ in eq.(\ref{UbarSP}) gives
 \beq \sum_{d'}\frac{1}{(2\pi)^3}\mid \ve{K}_{d'},d'\ra \bar{ \mathcal{U}}_{d',n':d,n} \la \ve{K}_{d},d \mid
 &&=\frac{1}{(2\pi)^3}\int \,d1\,d2\,d3\mid 1(a),3(b)\ra \la  1(a),2(b) \mid \ve{K}_d,d\ra \la \ve{K}_{d},d \mid  \eol &&\times  \la\la\Psi(n',\ve{x}=0)\mid \mathcal{V}(3, 2)
 \mid -\ve{K}_d, \psi_n \ra\ra
 \eol&&=\frac{1}{(2\pi)^3}\int \,d1\,d2\,d3\mid 1(a),3(b)\ra \la  1(a),2(b) \mid \ve{K}_d,d\ra \la \ve{K}_{d},d \mid \eol &&\times  \int d\ve{x}\exp(-\imath\ve{K}_d.\ve{x}) \la\la  \Psi(n',\ve{x}=0)\mid \mathcal{V}(3, 2)
 \mid  \Psi(n,\ve{x}) \ra\ra
 \eol&&=\frac{1}{(2\pi)^3}\int \,d1\,d2\,d3\mid 1(a),3(b)\ra \int d\ve{x}\la  1(a),2(b)\mid \exp(-\imath\bar{\ve{P}}.\ve{x})\mid \ve{K}_d,d\ra \la \ve{K}_{d},d \mid \eol &&\times   \la\la  \Psi(n',\ve{x}=0)\mid \mathcal{V}(3, 2)
 \mid  \Psi(n,\ve{x}) \ra\ra, \eol&&\label{UbarSP2}\eeq
where, in the last line, $\bar{\ve{P}}$ is the total momentum operator in the  sub-space of $\bar{B}$-space that involves nucleons $a$ and $b$.

The summation over $d$ can now be completed to give 
 \beq \sum_{d',d}\frac{1}{(2\pi)^3}\mid \ve{K}_{d'},d'\ra \bar{ \mathcal{U}}_{d',n':d,n} \la \ve{K}_{d},d \mid
 &&=\int \,d1\,d2\,d3\,d\ve{x}\mid 1(a),3(b)\ra  \la  1(a),2(b)\mid \exp(-\imath\bar{\ve{P}}.\ve{x}) \eol &&\times  \la\la  \Psi(n',\ve{x}=0)\mid \mathcal{V}(3, 2)\mid  \Psi(n,\ve{x}) \ra\ra. \eol&&\label{UbarSP3}\eeq

Hence
 \beq \bar{\mathcal{U}}_{n',n}&&=\int \,d1\,d2\,d3\,d\ve{x}\mid 1(a),3(b)\ra  \la  1(a),2(b)\mid \int d\ve{x}\exp(-\imath\bar{\ve{P}}.\ve{x}) \la\la  \Psi(n',\ve{x}=0)\mid \mathcal{V}(3, 2)\mid  \Psi(n,\ve{x}) \ra\ra \eol &&.\eol && \label{UbarSP4}\eeq

In a momentum basis for the one-nucleon states the operator $\bar{\ve{P}}$ can be replaced by the eigenvalue $\ve{k}_1+\ve{k}_2$ and the $\ve{x}$ integration gives
\beq \bar{\mathcal{U}}_{n',n}=\frac{1}{2}\int \,d1\,d2\,d3 \mid \ve{k}_1(a),\ve{k}_3(b)\ra  \la  \ve{k}_1(a),\ve{k}_2(b)\mid  \la\la  \Psi(n',\ve{x}=0)\mid \mathcal{V}(\ve{k}_3(b), \ve{k}_2(b))\mid -(\ve{k}_1+\ve{k}_2),\psi_n \ra\ra. \label{UbarSP51}\eeq

This is the correct form for the matrix element of a general momentum-conserving nucleon-$A$ interaction in the space of the three bodies $a+b+A$. The operator (\ref{UbarSP51}) is diagonal in the momentum and other quantum numbers of the non-interacting particle $a$. But note that the matrix element of $\mathcal{V}(\ve{k}_3(b), \ve{k}_2(b))$ on the right-hand-side does depend on $\ve{k}_1$, which might appear paradoxical as particle $a$ is not interacting with $A$. It is shown in Appendix \ref{gen2in3} that in fact this is exactly what is expected for a non-local interaction and finite target mass.

 The physical meaning of these formulas can also be seen by using a position basis for nucleons $a$ and $b$. In this case  $\exp(-\imath\bar{\ve{P}}.\ve{x})$ acts as a displacement operator so that
\beq \la \ve{r}_1(a),\ve{r}_2(b) \mid \exp(-\imath\bar{\ve{P}}.\ve{x})=  \la (\ve{r}_1-\ve{x}), (\ve{r}_2-\ve{x})  \mid .\label{Pdisp|}\eeq
In this case eq.(\ref{UbarSP4}) gives ( spin and iso-spin coordinates are omitted for clarity)
\beq \la \ve{r}_1'(a),\ve{r}_3'(b)\mid\bar{\mathcal{U}}_{n,n'}\mid \ve{r}_1''(a),\ve{r}_2''(b)\ra&&\!\!\!\!\!\!\!=\frac{1}{2}\int \,d1\,d2\,d3\,d\ve{x}\delta(\ve{r}_1'-\ve{r}_1)\delta(\ve{r}_3'-\ve{r}_3) \delta(\ve{r}_{1}''-\ve{r}_{1}+\ve{x})\delta(\ve{r}_{2}''-\ve{r}_{2}+\ve{x})\eol&&\!\!\!\!\!\!\! \times \la\la  \Psi(n,\ve{x}=0)\mid \mathcal{V}(\ve{r}_3, \ve{r}_2)\mid  \Psi(n',\ve{x}) \ra\ra\eol&&
\!\!\!\!\!\!\!=\frac{1}{2}\int \,d\ve{x}\,\delta(\ve{r}_{1}'-\ve{r}_{1}''-\ve{x})\la\la  \Psi(n,\ve{x}=0)\mid \mathcal{V}(\ve{r}_{3}'(b), \ve{r}_{2}''+\ve{x})(b)\mid  \Psi(n',\ve{x}) \ra\ra .\eol&& \label{Ubarcoord}\eeq

The structure of this formula reflects the fact that when  nucleon $b$ of the fictitious particle interacts with the target $A$ the position of the latter can change by $\ve{x}$ and this alters the position of $a$ with respect to the new overall centre-of-mass. Again it can be seen that for very large $A$ the $b-A$ scattering  associated with $\mathcal{V}$ in eq.(\ref{Ubarcoord}) is unlikely to generate a large shift in the position of the centre-of-mass of $A$. Values of $x$ close to zero will be strongly emphasised and the right-hand-side of eq.(\ref{Ubarcoord}) will become diagonal in the position of $a$.

Note that the the expressions for $\bar{\mathcal{U}}_{n',n}$ given in this section all use isospin notation. In the next subsection the same operator is expressed in a notation traditionally used for $n+p+A$ applications.  
 
\subsection{Isospin considerations.}
Practical calculations usually deal with couplings between target states of the same charge so that only terms with isospin $\tau_3=\tau_2$ contribute to $\bar{\mathcal{U}}_{n,n'}$ in eq.(\ref{UbarSP4}). For  $d$ states with $T=0$ as in the incident deuteron,  the relevant matrix element of $\bar{\mathcal{U}}_{n,n'}$  is $\la \Xi_{0,0}(a,b)\mid\bar{\mathcal{U}}_{n',n}\mid \Xi_{0,0}(a,b)\ra$ where
\beq \Xi_{0,0}(a,b)=\frac{1}{\sqrt{2}}(\chi_P(a)\chi_N(b)-\chi_N(a)\chi_P(b)).\label{xi00}\eeq
Carrying out the isospin calculation gives
\beq  \la\Xi_{0,0}(a,b)\mid\bar{\mathcal{U}}_{n',n}\mid \Xi_{0,0}(a,b)\ra=\bar{\mathcal{U}}^{PA}+\bar{\mathcal{U}}^{NA},\label{UbarSP42}\eeq\
where
\beq \bar{\mathcal{U}}^{PA}=&&\!\!\!\int \,d1'\,d2'\,d3'\,\mid 1'(P),3'(N)\ra  \la  1'(P),2'(N)\mid  \frac{1}{2}  \int d\ve{x}\exp(-\imath\bar{\ve{P}}.\ve{x}) \la\la  \Psi(n',\ve{x}=0)\mid \mathcal{V}(3'(N), 2'(N))\mid  \Psi(n,\ve{x}) \ra\ra, \eol&&\label{UbarPA}\eeq
and
\beq \bar{\mathcal{U}}^{NA}=&&\int \,d1'\,d2'\,d3'\,\mid 1'(N),3'(P)\ra  \la  1'(N),2'(P)\mid \frac{1}{2}  \int d\ve{x}\exp(-\imath\bar{\ve{P}}.\ve{x}) \la\la  \Psi(n',\ve{x}=0)\mid \mathcal{V}(3'(P), 2'(P))\mid  \Psi(n,\ve{x}) \ra\ra.\eol && \label{UbarNA}\eeq
The dashed variables $1',2',3'$ are space and intrinsic spin state-labels.
 
 For $d$ states of nucleons $a$ and $b$ with $T=0$, these expressions separate neutron and proton interactions with the target nucleons $A$, including the factor of $\frac{1}{2}$ that appears in eqs.(\ref{Uint}) and (\ref{xrepME}). A similar separation occurs in the $T=1,T_3=0$ channel.
 \subsection{Deuteron elastic and inelastic scattering, elastic and inelastic deuteron break-up.}
 Using the results of eqs.(\ref{UbarPA}) and (\ref{UbarNA}) the three-body model Hamiltonian $\bar{\mathcal{H}}$ in eq(\ref{HbartA}) and the coupled equations 
 (\ref{Beq}) can be written
  \beq (E+\imath \epsilon -\bar{T}-\bar{h}_{A}-\bar{h}_2-\bar{\mathcal{U}}^{PA}-\bar{\mathcal{U}}^{NA})\mid \Psi^{(\epsilon)}_{d_0,0}\ra=\imath \epsilon \mid d_0,0\ra.\label{Beq2}\eeq
  The solution to this equation is $\mid \Psi^{(\epsilon)}_{d_0,0}\ra$, a state in $\bar{B}$-space whose components form a set of overlap functions based on the exact scattering state corresponding to a deuteron incident on an $A$-nucleon target in its ground state. There is a component for all possible isospin zero states of the the 2-nucleon system and all states of the $A$-nucleon target. In the limit $\epsilon \rightarrow 0^+$ the only component with an incoming plane wave is the $d_0,0$ channel. All other channels have outgoing waves only.
    
  The techniques described in Sub-section \ref{FeshbachTheory} can be used to derive expressions for the effective interaction that generates a sub-space of components of $\mid \Psi^{(\epsilon)}_{d_0,0}\ra$. For example the effective interaction that generates the component in which the deuteron and target begin and end in their ground states, i.e., the deuteron optical potential, is given by (compare eqs.(\ref{Ueff}) and (\ref{Vhatopt}))
  \beq \la K_{d_0}\mid V_{d_0}\mid K'_{d_0}\ra=\la K_{d_0},d_0,0\mid \bar{U}^{(d_0,0)}_{\mathrm{eff}}\mid K'_{d_0} d_0,0\ra\label{dopt}\eeq 
  where $\bar{U}^{(d_0,0)}_{\mathrm{eff}}$ satisfies
   \beq \bar{U}^{(d_0,0)}_{\mathrm{eff}}=(\bar{\mathcal{U}}^{PA}+\bar{\mathcal{U}}^{NA})+(\bar{\mathcal{U}}^{PA}+\bar{\mathcal{U}}^{NA})\frac{1}{(E+\imath \epsilon-\bar{T}-\bar{h}_A-\bar{h}_2)}\bar{Q}^{d_0}_0\bar{U}^{(d_0,0)}_{\mathrm{eff}}, \label{Ubareff}\eeq
 and where $\bar{Q}^{d_0}_0$ projects onto all states $\mid d,n\ra$ with $d\neq d_0$, all $n$ and $d=d_0, n\neq 0$. In comparing with result (\ref{Vhatopt}) in the nucleon+$A$ case note that a modified notation is being used  where the  label $d$ refers to the intrinsic state of the nucleon pair and excludes their total momentum $\ve{K}_d$.
 \section{Application to 3-body models of $A(d,p)B$ reactions. }\label{multscatt}
Three-body models of the $n+p+A$ system are based on an effective interaction that differs from $\bar{U}^{(d_0,0)}_{\mathrm{eff}}$ of eq.(\ref{Ubareff}), \cite{TJ21}. The $\bar{Q}^{d_0}_0$ space includes elastic deuteron break-up, whereas 3-body deuteron stripping models treat these and their coupling to the  elastic deuteron channel explicitly. Accordingly, the correct effective interaction that describes the  couplings between these states that begin and end with the target in its ground state is $ \la d, 0\mid\bar{U}_{\mathrm{eff}}\mid d' 0\ra$ where $\bar{U}_{\mathrm{eff}}$ satisfies 
 \beq \bar{U}_{\mathrm{eff}}=(\bar{\mathcal{U}}^{PA}+\bar{\mathcal{U}}^{NA})+(\bar{\mathcal{U}}^{PA}+\bar{\mathcal{U}}^{NA})\frac{1}{(E+\imath \epsilon-\bar{T}-\bar{h}_A-\bar{h}_2)}\bar{Q}_0\bar{U}_{\mathrm{eff}}, \label{Ubareffd0}\eeq
and where $\bar{Q}_0$ projects onto  states $\mid d,n\ra$ which include all $d$ states and target states with $n\neq 0$. By construction, $\la d, 0\mid\bar{U}^{(d,0)}_{\mathrm{eff}}\mid d',0\ra$ is an operator in the space of the two nucleons in $d$ space and  in the space of degrees of freedom associated with the target intrinsic ground state. This description is readily generalised to treat a finite number of excited target states on the same  footing as the ground state and exclude them from $\bar{Q}^d_0$. 
 
 The operators $\bar{\mathcal{U}}^{PA}$ and $\bar{\mathcal{U}}^{NA}$  each involve the interaction of one of the nucleons in the $d$ pair with the target nucleons. As such they describe the same physical processes as the interaction $\hat{ \mathcal{U}}$ defined in eq.(\ref{Uint}) and appears in the expression (\ref{Vhatopt}) for the nucleon+$A$ optical potential, but here written in a form appropriate for a non-local, translationally and rotationally invariant interaction in the 3-body context.  
  \subsection{Relation of the 3-body effective interaction $\bar{U}^{(d,0)}_{\mathrm{eff}}$ to  nucleon optical potentials.}
  It is straight forward to separate out the $PA$ and {NA} components together with their the target excitation contributions to $\bar{U}_{\mathrm{eff}}$, eq.(\ref{Ubareffd0}) using standard multiple scattering manipulations.
  
  Following \cite{Tim14}, Appendix 2, the operators $\bar{V}^{PA}$ and $\bar{V}^{NA}$ are defined by 
  \beq \bar{V}^{PA}=&&\bar{\mathcal{U}}^{PA}+\bar{\mathcal{U}}^{PA}\bar{Q}_{0}\bar{G}_0\bar{V}^{PA},\eol
 \bar{V}^{NA}=&&\bar{\mathcal{U}}^{NA}+\bar{\mathcal{U}}^{NA}\bar{Q}_{0}\bar{G}_0\bar{V}^{NA}.\label{VPVN}\eeq 
 where                                                                                                                                                                                                                                                                                                                                                                                                                                                                                                                                                                       \beq \bar{G}_0=\frac{1}{(E+\imath \epsilon-\bar{T}-\bar{h}_A-\bar{h}_2)}.\label{Gbar0}\eeq
 The contributions of  $\bar{V}^{PA}$ and $\bar{V}^{NA}$ to $\bar{U}_{\mathrm{eff}}$ are most simply obtained by first defining the subsidiary operators $\bar{W}^{NA}$ and $\bar{W}^{PA}$ through the coupled equations         
\beq \bar{W}^{PA}=&&\bar{V}^{PA}+\bar{V}^{PA}\bar{Q}_{0}\bar{G}_0\bar{W}^{NA},\eol
 \bar{W}^{NA}=&&\bar{V}^{NA}+\bar{V}^{NA}\bar{Q}_{0}\bar{G}_0\bar{W}^{PA}.\label{WPWN}\eeq
The definitions (\ref{VPVN}) give
\beq (1-\bar{\mathcal{U}}^{PA}\bar{Q}_{0}\bar{G}_0)\bar{V}^{PA}=&&\bar{\mathcal{U}}^{PA},\eol
(1-\bar{\mathcal{U}}^{NA}\bar{Q}_{0}\bar{G}_0) \bar{V}^{NA}=&&\bar{\mathcal{U}}^{NA}.\label{VPVN2}\eeq 
Multiplying the first and second equations in eq.(\ref{WPWN}) by, respectively, $1-\bar{\mathcal{U}}^{PA}\bar{Q}_{0}\bar{G}_0$ and $1-\bar{\mathcal{U}}^{NA}\bar{Q}_{0}\bar{G}_0$, adding the results and using (\ref{VPVN2}) gives
 \beq (1-\bar{\mathcal{U}}^{PA}\bar{Q}_{0}\bar{G}_0)\bar{W}^{PA}+(1-\bar{\mathcal{U}}^{NA}\bar{Q}_{0}\bar{G}_0)\bar{W}^{NA}&&\!\!\!\!\!\!\!=(\bar{\mathcal{U}}^{PA}+\bar{\mathcal{U}}^{NA})+\bar{\mathcal{U}}^{PA}\bar{Q}_{0}\bar{G}_0\bar{W}^{NA}+\bar{\mathcal{U}}^{NA}\bar{Q}_{0}\bar{G}_0\bar{W}^{PA}\eol&&
 \!\!\!\!\!\!\!=(\bar{\mathcal{U}}^{PA}+\bar{\mathcal{U}}^{NA})+\bar{\mathcal{U}}^{PA}\bar{Q}_{0}\bar{G}_0(\bar{W}^{NA}+\bar{W}^{PA})+\bar{\mathcal{U}}^{NA}\bar{Q}_{0}\bar{G}_0(\bar{W}^{PA}+\bar{W}^{NA})\eol &&
-\bar{\mathcal{U}}^{PA}\bar{Q}_{0}\bar{G}_0\bar{W}^{PA}-\bar{\mathcal{U}}^{NA}\bar{Q}_{0}\bar{G}_0\bar{W}^{NA}. \label{WPWN2}\eeq
Hence
\beq (\bar{W}^{NA}+\bar{W}^{PA})=(\bar{\mathcal{U}}^{PA}+\bar{\mathcal{U}}^{NA})+(\bar{\mathcal{U}}^{PA}+\bar{\mathcal{U}}^{NA})\bar{G}_0\bar{Q}_0(\bar{W}^{NA}+\bar{W}^{PA}), \label{Ubareffd02}\eeq 
and a comparison with eq.(\ref{Ubareffd0}) shows that
\beq \bar{U}_{\mathrm{eff}}=(\bar{W}^{PA}+\bar{W}^{NA}). \label{WPWN3}\eeq
Eqs.(\ref{WPWN3}), (\ref{WPWN}) and (\ref{VPVN})  provide a method for expressing $\bar{U}_{\mathrm{eff}}$ in terms of the separate contributions from  $\bar{\mathcal{U}}^{PA}$ and $\bar{\mathcal{U}}^{NA}$. 

The operators $\bar{V}^{PA}$ and $\bar{V}^{NA}$ show a strong resemblance to the operator $\hat{U}_{\mathrm{eff}}$ defined in eq.(\ref{Ueff2}) whose target ground state expectation value gives the nucleon optical potential.  However, the Green's function, $\bar{G}_0$, that appears in  eqs.(\ref{WPWN3}), (\ref{WPWN}), and (\ref{VPVN}) involve both nucleons $N$ and $P$ and their interaction $V_{NP}$ rather than just the kinetic energy operator associated with the single incident nucleon that appears in the Green's function in eq.(\ref{Ueff2}). Early explorations of 3-body effects in $A(d,p)B$ reactions ignored these considerations and simply replaced $\bar{V}^{PA}$ and $\bar{V}^{NA}$ by phenomenological optical potentials corresponding to nucleons with half the incident deuteron c.m. energy. Approximate ways of handling this situation and their quantitative effects of corrections to standard procedures in the context of the ADWA method for $(d,p)$ reactions are reviewed in detail in \cite{TJ21}. An extension of the same ideas that takes into account three-body terms in $\bar{U}_{\mathrm{eff}}$ arising from the second terms on the right in eqs.(\ref{WPWN}) and correspond  to target excitations by $N$ followed by de-excitation by $P$, and vice versa, are explained and applied to some specific $(d,p)$ reactions in \cite{Din19}. 
   \section{Conclusions.}\label{conclude}
It has been shown that the set of overlap functions, $\la\la \Psi(n,\ve{x}=0\mid A_{d}\mid \Psi^{\epsilon}_{d_0,0}\ra\ra$,  associated with the antisymmetrised, translationally invariant  centre-of-mass wave function, $\mid \Psi^{\epsilon}_{d_0,0}\ra\ra$, corresponding to a scattering state in which a plane wave deuteron is incident on an $A$-nucleon target in its ground state, satisfy coupled equations, eq. (\ref{dAscatt6}), in a 3-body model in which a fictitious particle with  the reduced mass of the deuteron that can exist in any one of a set of states corresponding to the spectrum of the real 2-nucleon system and interacts with an A-nucleon system with the excitation spectrum of the real target..  States of the fictitious particle and the target are coupled by  an interaction generated by the nucleon-nucleon interaction including exchange terms, eq.(\ref{Ubardnd'n'}). The physical meaning of this interaction is discussed and its relation to to the interaction that generates the nucleon optical potential is elucidated. The structure of the theory suggests that it might be possible to usefully generalise the treatment here to other systems for which a few-body model is appropriate.  

   In the application of 3-body models to $A(d,pB)$ reactions it is frequently the case that only the component of the many body scattering wave function with the target in its ground state is explicitly taken in to account in the calculation of the reaction matrix element, but  deuteron  elastic and break-up components are explicitly taken into account using approximate or exact treatments of the 3-body dynamics\cite{TJ21}. It has been shown here how to construct the effective 3-body interaction, including exchange effects, that should be used to calculate the overlap functions associated with this component, including the effects of target excitation. Note however that because of the non-orthogonality of antisymmetrised channel states, to reproduce the fully antisymmetrised many-nucleon scattering wave function requires a further step using a matrix of target transition densities. However, for the purpose of formulating a theory of the nucleon optical potential and relating it to the effective interaction used in standard 3-body models of the $A(d,p)B$ reaction, this step is not necessary. The analysis in this provides a justification for the models used in \cite{Tim14} and \cite{Din19} which were based on a theory that was not fully antisymmetrised.
   \section{Acknowledgements}
   In the course of formulating the approach followed in this paper I have enjoyed a fruitful correspondence with M.C.Birse. Support from the UK Science and Technology Facilities Council through the grant STFC ST/000051/1 is acknowledged.

\appendix     
\section{The factor $\frac{1}{2}$ in the definition of $\hat{\mathcal{U}}$ for $A=1$.}\label{A}
In the case $A=1$, with both the beam and target particles neutrons, the general expression expression for $\hat{U}$ is
  \beq 
 \la \ve{k},n\mid\hat{\mathcal{U}}\mid \ve{k}',n'\ra&=&\frac{1}{2}\int\,d\ve{k}_3\,d\ve{k}_4\la \ve{k},\ve{k}_3\mid V_{\mathcal{A}}\mid \ve{k}',\ve{k}_4\ra\la\la\Psi_n,\ve{x}=0\mid a^\dagger_{\ve{k}_3}a_{\ve{k}_4}\mid  -\ve{k}', \psi_{n'}\ra\ra.\label{Hhat22}\eeq
 If $A=1$ indices $n,n'$ become the neutron spin projections $\sigma =\pm\frac{1}{2}$ and the kets $\mid \Psi_n,\ve{x}=0\ra\ra$ and $\mid  -\ve{k}', \psi_{n'}\ra\ra$ become
 \beq \mid  -\ve{k}', \psi_{n'}\ra\ra &&\rightarrow (2\pi)^{(3/2)}a^\dagger_{-\ve{k}',\sigma_{n'}}\mid 0\ra\ra\eol
 \mid \Psi_n,\ve{x}=0\ra&&\rightarrow \psi^\dagger(\ve{r}=0,\sigma_n)\mid 0\ra\ra. \label{A1}\eeq
Using 
 \beq [a_{\ve{k}},\psi^\dagger(\ve{r})]=\frac{1}{(2\pi)^{3/2}}\exp(-\imath \ve{r}.\ve{k}).\label{akpsidaggercomm2}\eeq
   the matrix element in eq.(\ref{Hhat22}) becomes
  \beq \la\la\Psi(n,\ve{x}=0)\mid a^\dagger_{\ve{k}_3}a_{\ve{k}_4}\mid  -\ve{k}', \psi_{n'}\ra\ra\rightarrow \delta(\ve{k}_4+\ve{k}')\delta_{\sigma_4,\sigma_{n'}}\delta_{\sigma_3,\sigma_n},\label{A12}\eeq
  and eq.(\ref{Hhat22}) reduces to
   \beq 
 \la \ve{k},\sigma;\sigma_n\mid\hat{\mathcal{U}}\mid \ve{k}',\sigma';\sigma_{n'}\ra&&=\frac{1}{2}\int\,d\ve{k}_3\,\la \ve{k},\sigma;\ve{k}_3,\sigma_n\mid V_{\mathcal{A}}\mid \ve{k}'\sigma';-\ve{k}',\sigma_{n'}\ra\eol
 &&=\frac{1}{2}\int\,d\ve{k}_3\,(\la \ve{k},\sigma(1);\ve{k}_3,\sigma_n(2)\mid V\mid \ve{k}',\sigma'(1);-\ve{k}',\sigma_{n'}(2)\ra\eol
 &&\,\,\,\,\,\,\,\,\,\,\,\,\,\,\,\,\,\,\,\,\,\,\,\,\,\,\,\,\,\,\,\,\,\,\,\,\,\,\,\,\,\,\,\,\,\,\,\,\,\,\,\,\,\,\,\,-\la \ve{k},\sigma(1);\ve{k}_3,\sigma_n(2)\mid V\mid \ve{k}',\sigma'(2);-\ve{k}',\sigma_{n'}(1)\ra )\eol
 .\label{A13}\eeq
 For example, with $V$ a local, translationally and rotationally invariant spin independent interaction, $V(\mid \ve{r}_1-\ve{r}_2\mid)$, (\ref{A13}) reduces to
 \beq 
 \la \ve{k},\sigma;\sigma_n\mid\hat{\mathcal{U}}\mid \ve{k}',\sigma';\sigma_{n'}\ra&&=\frac{1}{2}\int\,d\ve{k}_3\,\la \ve{k},\sigma;\ve{k}_3,\sigma_n\mid V_{\mathcal{A}}\mid \ve{k}'\sigma';-\ve{k}',\sigma_{n'}\ra\eol
 &&=\frac{1}{2}\int\,d\ve{k}_3\,(\la \ve{k}(1);\ve{k}_3(2)\mid V\mid \ve{k}'(1);-\ve{k}'(2)\ra\delta_{\sigma,\sigma'}\delta_{\sigma_{n},\sigma_{n'}}\eol
 &&\,\,\,\,\,\,\,\,\,\,\,\,\,\,\,\,\,\,\,\,\,\,\,\,\,\,\,\,\,\,\,\,\,\,\,\,\,\,\,\,\,\,\,\,\,\,\,\,\,\,\,\,\,\,\,\,-\la \ve{k}(1);\ve{k}_3(2)\mid V\mid \ve{k}'(2);-\ve{k}'(1)\ra \delta_{\sigma,\sigma_{n'}}\delta_{\sigma_{n},\sigma'}).\label{A14}\eeq
 In a channel spin representation this can be written
 \beq 
 \la \ve{k},S\Sigma\mid\hat{\mathcal{U}}\mid \ve{k}',S'\Sigma'\ra &&=\delta_{S,S'}\delta_{\Sigma,\Sigma'}\int\,d\ve{k}_3\,\frac{1}{2}(\la \ve{k}(1);\ve{k}_3(2)\mid V\mid \ve{k}'(1);-\ve{k}'(2)\ra+(-)^S\la \ve{k}(1);\ve{k}_3(2)\mid V\mid \ve{k}'(2);-\ve{k}'(1)\ra .\eol&&\label{A14}\eeq
The factor $\frac{1}{2}$ is exactly what is needed to guarantee that even partial waves for $S=0$ and odd partial waves for $S=1$ contribute with the correct weight.

\section{The inhomogeneous term $\imath \epsilon\hat{\mathcal{K}}$, eq.(\ref{Beqcoupled}), in the nucleon$+A$ case.}\label{B}
The matrix elements of $\hat{\mathcal{K}}$ are defined in eq.(\ref{Kbarnkoko}) as
\beq \hat{\mathcal{K}}_{n,\ve{x}=0;0,\ve{k},\ve{k}_0}&&=\la\la\Psi_n,\ve{x}=0 \mid a_{\ve{k}}a^\dagger_{\ve{k}_0}\mid -\ve{k}_0,\psi_0 \ra\ra \eol
&&=\delta(\ve{k}-\ve{k}_0)\delta_{n,0}-\mathcal{K}_{n,\ve{x}=0;0,\ve{k},\ve{k}_0}
 ,\label{Kbarnkoko2}\eeq
where 
\beq \mathcal{K}_{n,\ve{x}=0;0,\ve{k},\ve{k}_0}=\la\la\Psi(n,\ve{x}=0 )\mid a^\dagger_{\ve{k}_0} a_{\ve{k}} \mid -\ve{k}_0,\psi_0 \ra\ra \label{Knkko}\eeq
The contribution to the solution of eqs.(\ref{Beqcoupled}) from the two terms on the right in eq.(\ref{Kbarnkoko2}) can be elucidated by using  the identity

 \beq (E^+-\hat{\mathcal{H}})^{-1}=(E^+-\hat{\mathcal{H}}_0)^{-1}+(E^+-\hat{\mathcal{H}})^{-1}\hat{\mathcal{U}}(E^+-\hat{\mathcal{H}}_0)^{-1}, \label{identity1}\eeq
    so that
\beq (E+\imath \epsilon-\hat{\mathcal{H}})^{-1}\imath \epsilon \hat{\mathcal{K}}(2\pi)^{3/2}\mid 0,\ve{k}_0\ra=
(1+(E^+-\hat{\mathcal{H}})^{-1}\hat{\mathcal{U}})(E^+-\hat{\mathcal{H}}_0)^{-1}\imath \epsilon\hat{\mathcal{K}}\mid \ve{k}_0,0\ra. \label{identity2}\eeq
The first term in eq.(\ref{Kbarnkoko2}) gives
\beq (1+(E^+-\hat{\mathcal{H}})^{-1}\hat{\mathcal{U}})\mid \ve{k}_0,0\ra, \label{Kbarnkoko3}\eeq 
which, for $\epsilon \rightarrow 0+$, is just the standard scattering state solution of the coupled equations with a plane wave in the incident channel plus outgoing spherical waves asymptotically in all channels.

The contribution from the second term on the right in eq.(\ref{Kbarnkoko2}) requires explicit forms for the bra and ket in eq.(\ref{Knkko}). These are
\beq \mid -\ve{k}_0,\psi_0 \ra\ra=\frac{1}{\sqrt{A!}}\int d\ve{r}_1\dots\ve{r}_A\exp(-\imath \ve{k}_0.\ve{R}_A) \psi_0(\ve{r}_1,\dots, \ve{r}_A)
 \psi^\dagger(\ve{r}_A)\dots \psi^\dagger(\ve{r}_1)\mid 0 \rangle \rangle\ \label{psirPsi03}\eeq
and
\beq \mid \Psi(n,\ve{x}=0) \ra\ra=\frac{1}{\sqrt{A!}}\int d\ve{r}_1\dots\ve{r}_A\delta(\ve{R}_A) \psi_n(\ve{r}_1,\dots,\ve{r}_A)\psi^\dagger(\ve{r}_A)\dots \psi^\dagger(\ve{r}_1)\mid 0 \rangle \rangle\label{psirPsin3}\eeq 
To proceed further, it is convenient to transform to  the set of $A-1$ translationally invariant Jacobi coordinates $\ve{\chi}_1,\dots\,\ve{\chi}_{(A-1)}$  that together with the $A$-nucleon c.m., $\ve{R}_A$, are equivalent to the vectors $\ve{r}_1, \dots \ve{r}_A$ 
and have a transformation Jacobian equal to  $+1$. They are defined so that 
\beq\ve{\chi}_{j-1}= \ve{r}_{j}-\ve{R}_{(j-1)},\,\,\ve{R}_{(j-1)}=(\sum_{i=1}^{i=j-1}\ve{r}_i)/(j-1),\,\,\,j=2,\dots,A.\label{chijdef} \eeq

Inserting expressions (\ref{psirPsi03}) and (\ref{psirPsin3}) into (\ref{Knkko}) and using the result (\ref{akpsidaggercomm2}) gives
\beq\la\la\Psi(n,\ve{x}=0)\mid\!\!\!\!\!\!\!&& a^\dagger _{\ve{k}_0} a _{\ve{k}}\mid  -\ve{k}_0,  \psi_{0} \ra\ra=\frac{A}{(2\pi )^3}\int d\ve{\chi}_{A-1}'\,d\ve{\chi}_{A-1}\eol&&\times \exp(\imath( \ve{k}_0+\frac{\ve{k}}{A}).\ve{\chi}_{A-1}')\,\exp(-\imath(\ve{k}+\frac{\ve{k}_0}{A}).\ve{\chi}_{A-1})\eol&&\times\int d\ve{\chi}_{1}\, \dots d\ve{\chi}_{(A-2)}\, \psi^*_n(\ve{\chi}_1,\dots\,\ve{\chi}_{(A-2)},\ve{\chi}_{A-1}')\psi_0(\ve{\chi}_1,\dots\,\ve{\chi}_{(A-2)},\ve{\chi}_{A-1}),\eol&&\label{xibasisK2}\eeq 
The significance of the momentum $(\ve{k}+\frac{\ve{k}_0}{A})$ that appears in eq.(\ref{xibasisK2}) is that $(\frac{\hbar\ve{k}}{m}+\frac{\hbar\ve{k}_0}{Am})$ is the relative velocity of a nucleon with momentum $\ve{k}$ and nucleus of mass $Am$ with momentum $-\ve{k}_0$. Similarly, $(\frac{\hbar\ve{k}_0}{m}+\frac{\hbar\ve{k}}{Am})$ is the relative velocity of a nucleon with momentum $\ve{k}_0$ and nucleus of mass $Am$ with momentum $-\ve{k}$. These are the velocities associated with the Jacobi coordinate $\ve{\chi}_{A-1}$ that is defined as the  distance between nucleon $A$ and the c.m. of the $(A-1)$ nucleons $\ve{r}_1,\dots, \ve{r}_{(A-1)}$. They are the velocities that are relevant to the evaluation of the kets $a _{\ve{k}}\mid  -\ve{k}_0,  \psi_{0} \ra\ra$ and $a _{\ve{k}_0}\mid  -\ve{k},  \psi_{n} \ra\ra$ that are overlapped on the left-hand-side of eq.(\ref{xibasisK2}). 

In the subsequent discussion it is convenient to write the result (\ref{xibasisK2}) in the form
\beq\mathcal{K}_{n,\ve{x}=0;0,\ve{k},\ve{k}_0}=\int d\ve{\chi}'_{(A-1)} \exp(\imath (\ve{k}_0+\frac{\ve{k}}{A}).\ve{\chi}'_{(A-1)})F_{n,0}(\ve{\chi}'_{(A-1)}), \label{DM1252}\eeq
 where
 \beq F_{n,0}(\ve{\chi}'_{(A-1)})&&=\int d\ve{\chi}_{1}\, \dots d\ve{\chi}_{(A-1)} \psi^*_n(\ve{\chi}_1,\dots\,\ve{\chi}_{(A-2)},\ve{\chi}'_{(A-1)})\eol
&&\times \exp(-\imath (\ve{k}+\frac{\ve{k}_0}{A}).\ve{\chi}_{(A-1)})\psi_0(\ve{\chi}_1,\dots\,\ve{\chi}_{(A-2)},\ve{\chi}_{(A-1)})\label{Fn0}\eeq

If  $\psi_0$ and  $\psi_n$ are both bound states the overlap $F_{n,0}(\ve{\chi}'_{(A-1)})$ will decay exponentially for large $\ve{\chi}'_{(A-1)}$. Eq.(\ref{DM1252}) will then define a square integrable function of the momentum $(\ve{k}+\frac{\ve{k}_0}{A})$ and $(E+\imath \epsilon-\hat{\mathcal{H}}_0)^{-1} \mathcal{K}\mid \ve{k}_0,0\ra$ will have a finite limit when $\epsilon \rightarrow 0$. As a result $\imath \epsilon \frac{1}{E+\imath \epsilon-\hat{\mathcal{H}}_0)} \mathcal{K}\mid \ve{k}_0,0\ra$ will vanish in the same limit and give no contribution to the solution of the coupled equations (\ref{scattwf4}).

It will be assumed that the target ground state $\psi_0$ is bound but the situation is more complicated when $\psi_n$ is a continuum state. In this case the overlap $F_{n,0}(\ve{\chi}'_{(A-1)})$ will oscillate at large $\ve{\chi}'_{(A-1)}$ with wave numbers determined by energy conservation.  This situation can be modelled by assuming 
\beq F_{n,0}(\ve{\chi}'_{(A-1)})=\exp(-\imath \ve{k}_n.\ve{\chi}'_{(A-1)})C_{n0}, \label{Fn02}\eeq
so that
\beq\mathcal{K}_{n,\ve{x}=0;0,\ve{k},\ve{k}_0}&&=\int d\ve{\chi}'_{(A-1)} \exp(\imath (\ve{k}_0+\frac{\ve{k}}{A}).\ve{\chi}'_{(A-1)})\exp(-\imath \ve{k}_n.\ve{\chi}'_{(A-1)})C_{n0}\eol
   &&=(2\pi)^3C_{n0}\delta((\ve{k}_0+\frac{\ve{k}}{A})-\ve{k}_n).\label{singcase1}\eeq
   
   The meaning of the Jacobi coordinate  $\ve{\chi}_{(A-1)}$ as the separation of nucleon $A$ from the c.m. of the $(A-1)$ nucleons $1,\dots ,(A-1)$ in an intrinsic state with energy $E^{(A-1)}$ means that $\ve{k}_n$ must be related to $E_n$ and $E^{(A-1)}$ by
   \beq E_n=\frac{A}{(A-1)}\frac{\hbar^2k_n^2}{2m} +E^{(A-1)}.\label{EnA} \eeq
 Using eq.(\ref{singcase1}) gives
\beq  \frac{\imath \epsilon}{(E^+-\hat{\mathcal{H}}_0)}\mathcal{K}\mid 0,\ve{k}_0\ra =\sum_n\int d\ve{k} \frac{\imath \epsilon}{(\frac{(A+1)}{A}\frac{\hbar^2k_0^2}{2m}-\epsilon_0-(E_n+\frac{(A+1)}{A}\epsilon_k)+\imath \epsilon)}\delta (\ve{k}_n-(\frac{\ve{k}}{A}+\ve{k}_0))(2\pi)^3C_{n0}.\label{G0KA}\eeq
Using the delta function in the numerator, the denominator in (\ref{G0KA}) can be written
\beq (\frac{(A+1)}{A}\frac{\hbar^2k_0^2}{2m}-\epsilon_0-(E_n+\frac{(A+1)}{A}\epsilon_k)+\imath \epsilon)&&=(\frac{(A+1)}{A}\frac{\hbar^2k_0^2}{2m}-\epsilon_0-(\frac{A}{(A-1)}\frac{\hbar^2k_n^2}{2m} +E^{(A-1)}+\frac{(A+1)}{A}\epsilon_k)+\imath \epsilon)\eol
&&=(-\frac{A}{(A-1)}\frac{\hbar^2(\frac{\ve{k}_0}{A}+\ve{k})^2}{2m}-\epsilon_0-E^{(A-1)}+\imath \epsilon).\label{denA1}\eeq 
For a stable ground state $\psi_0$
\beq -\epsilon_0<E^{(A-1)},\label{stabiitycond}\eeq 
for all $(A-1)$-nucleon states. The denominator (\ref{denA1}) never vanishes and so
\beq \lim_{\epsilon\rightarrow 0}\frac{\imath \epsilon}{(E^+-\hat{\mathcal{H}}_0)}\mathcal{K}\mid 0,\ve{k}_0\ra=0.\label{limden}\eeq
and the coupled equations (\ref{scattwf4}) can be replaced by
 \beq \sum_{1,n'}((E^+-E_n-\frac{(A+1)}{A}\epsilon_k)\delta(\ve{k}-\ve{k}_1)\delta_{n,n'}-\frac{1}{2}\la\la\Psi(n,\ve{x}=0)\mid\mathcal{V}_{\ve{k},\ve{k}_1}\mid -\ve{k}_1. \psi_{n'}\ra\ra\la\la\Psi(n',\ve{x}=0) \mid a_{\ve{k}_1}\mid \Psi_{\ve{k}_0,0}^{(\epsilon)}\ra\ra \eol
=\imath \epsilon (2\pi)^{3/2}\delta(\ve{k}-\ve{k}_0)\delta_{n,0}. \label{scattwf42}\eeq

  \section{The commutator $ [A_{d'},A^\dagger_{d}]_-$.}\label{C}
The definition of $A^\dagger_{d}$  in eq.(\ref{Adaggerdgen4}) make it clear that the basic result needed for the commutator $[A_{d'} ,A_{d}^\dagger ]_-$ is the commutator $[a_4 a_3,a_1^\dagger a_2^\dagger]_-$. Using the fermion anticommutation relations gives 
\beq  [a_4 a_3,a_1^\dagger a_2^\dagger ]_-=\delta_{1,3}\delta_{2,4} -\delta_{1,4}\delta_{2,3}+\delta_{1,4}a_2^\dagger a_3-\delta_{1,3}a_2^\dagger a_4+\delta_{2,3}a_1^\dagger a_4-\delta_{2,4}a_1^\dagger a_3.\label{R2}\eeq
The last four terms all give the same contribution when account is taken of the antisymmetry of the kets $\mid d(a,b)\ra$, and so
\beq  [A_{d'},A^\dagger_{d}]_- &&=\frac{1}{2}\int d\,1 \d\,2\, d\,3\, d\,4 ( \la 3(a),4(b) \mid d'(a,b)\ra^* \la 1(a),2(b) \mid d(a,b)\ra  [a^\dagger_1a^\dagger_2, a_4a_3]_- \eol&&
=\delta_{d',d}-2\int d\,1 d\,2\, d\,3   \la 1(a),3(b) \mid d'(a,b)\ra^*\la 1(a),2(b) \mid d(a,b)\ra a^\dagger_2a_3
.\label{deutcomms3}\eeq

 Note that, if the antisymmetrised 2-nucleon states $\mid 1,2\ra$ defined in eq.(\ref{ijstate}) are used the factor 2 in eq.(\ref{deutcomms3}) disappears.
 \beq  [A_{d'},A^\dagger_{d}]_- =\delta_{d',d}-\int d\,1 d\,2\, d\,4   \la 1,4 \mid d'(a,b)\ra ^*\la 1,2 \mid d(a,b)\ra a^\dagger_2a_4
.\label{deutcomms4}\eeq
\section{The current operator $J^\dagger_{d}$.} \label{D}
In the approach to scattering theory used by Villars \cite{Villars1967} the commutator $[H,A^\dagger_{d}]_-$ and and the current operator $J^\dagger_{d}$ play a crucial role. Here, an explicit expression is deduced for the quantity $J^\dagger_d$ defined by 
\beq [H,A^\dagger_{d}]_-=E_{d}A^\dagger_{d}+J^\dagger_{d}.\label{JKSdef}\eeq
where $H=T+V$ is the many-nucleon Hamiltonian in Fock space. The energy $E_{d}$ is defined using the conventions explained following eq.(\ref{phdeq}).
\beq E_{d}=\frac{\hbar^2K^2}{4m}+\epsilon_{d}. \label{EKddef}\eeq

 A general creation operator is written
\beq A^\dagger_{d}=\frac{1}{\sqrt{2}}\int d\,1 d\,2 \la 1(a),2(b) \mid d(a,b)\ra a^\dagger_1a^\dagger_2.
\label{Adaggerdgen42}\eeq
Here $\mid 1(a),2(b) \ra$ denotes one of the  two-nucleon states
\beq \mid 1(a),2(b)\ra=\mid 1(a)\ra \mid 2(b)\ra, \label{122} \eeq
and, following eq.(\ref{phdeq}), the function $\la 1(a),2(b)\mid d(a,b)\ra$ satisfies
\beq \la 1(a),2(b)\mid (E_{d}-T_1-T_2)\mid d(a,b)\ra=\int d3\,d4\,\frac{1}{2}\la 1,2\mid V_{\cal{A}}\mid 3,4\ra\la 3(a),4(b)\mid d(a,b)\ra, \label{phdeq2}\eeq
where $T_1$ and $T_2$ are nucleon kinetic energies. The notation of \cite{Johnson19}, Eq.(41), is used for the antisymmetrised matrix elements of the nucleon-nucleon 2-body interaction $V$.

From eq.(\ref{Adaggerdgen42})
\beq [H,A^\dagger_d]_-=\frac{1}{\sqrt{2}}\int d\,1 d\,2\la 1(a),2(b) \mid d(a,b)\ra [H,a^\dagger_1a^\dagger_2]_-.\label{HAcomm}\eeq
The kinetic energy part of $H$ gives
\beq [T,a^\dagger_1a^\dagger_2]_-= a^\dagger_1[T,a^\dagger_2]_-+[T,a^\dagger_1]_-a^\dagger_2=(T_1+T_2)a^\dagger_1a^\dagger_2,\label{TAcomm}\eeq
and the potential energy term is
\beq [V,a^\dagger_1a^\dagger_2]_-= a^\dagger_1[V,a^\dagger_2]_-+[V,a^\dagger_1]_-a^\dagger_2.\label{Acomm}\eeq
There are two distinct contributions from the potential energy commutators to $[H,A^\dagger_d]_-$.  One term includes the binding contributions to the deuteron ground state as in eq.(\ref{phdeq2}). The remaining terms describe the interaction between the nucleons in the deuteron and any other nucleons and vanish when they act on the vacuum. Next it will be explained how this separation can be made explicit.

 In the notation of \cite{Johnson19},eq.(46),
 \beq [V,a^\dagger_i]_-&&=\frac{1}{2}\int \,d3\, a^\dagger_{3}  \mathcal{V}(3, i), 
   \label{Vadaggercomm}\eeq
   where
 \beq \mathcal{V}(3, i)&&=
\int \,d4\, d5\,\la 3,4\mid V_{\mathcal{A}}\mid i,5\ra a^\dagger_{4}a_{5}.
\label{mathcalV}\eeq
The second term on the right of eq.(\ref{Acomm}) gives
\beq [V,a^\dagger_1]_-a^\dagger_2&&=\frac{1}{2}\int \,d3\,d4\, d5\,\la 3,4\mid V_{\mathcal{A}}\mid 1,5\ra a^\dagger_{3}a^\dagger_{4} a_5a^\dagger_{2} \eol &&=\frac{1}{2}\int \,d3\,d4\,\la 3,4\mid V_{\mathcal{A}}\mid 1,2\ra a^\dagger_{3}a^\dagger_{4} -\frac{1}{2}\int \,d3\,d4\, d5\,\la 3,4\mid V_{\mathcal{A}}\mid 1,5\ra a^\dagger_{3}a^\dagger_{4} a^\dagger_{2}a_{5}\eol&&=\frac{1}{2}\int \,d3\,d4\,\la 3,4\mid V_{\mathcal{A}}\mid 1,2\ra a^\dagger_{3}a^\dagger_{4} -\frac{1}{2}a^\dagger_{2}\int \,d3\,a^\dagger_{3}\mathcal{V}(3, 1), 
   \label{Vadaggercomm2}\eeq
   and the first term is
  \beq a^\dagger_1 [V,a^\dagger_2]_-&&=\frac{1}{2}\int \,d3\,d4\, d5\,\la 3,4\mid V_{\mathcal{A}}\mid 2,5\ra a^\dagger_{1}a^\dagger_{3}a^\dagger_{4} a_5 \eol &&=\frac{1}{2}a^\dagger_{1}\int \,d3\,a^\dagger_{3}\mathcal{V}(3, 2). 
   \label{Vadaggercomm3}\eeq
   
Putting the results (\ref{Vadaggercomm2}) and (\ref{Vadaggercomm3}) into (\ref{Acomm}) and using (\ref{TAcomm}) and the antisymmetry of the deuteron wave function $\la 1(a),2(b) \mid d(a,b)\ra$ gives
\beq [V,A^\dagger_d]_-&&=\frac{1}{\sqrt{2}}\int d\,1 d\,2 \la 1(a),2(b) \mid d(a,b)\ra [V,a^\dagger_1a^\dagger_2]_-\eol&&
=\int d\,1 d\,2\,\frac{1}{\sqrt{2}} \la 1(a),2(b) \mid d(a,b)\ra(a^\dagger_1[V,a^\dagger_2]_-+[V,a^\dagger_1]_-a^\dagger_2)\eol&&
=\int d\,1 d\,2\,\frac{1}{\sqrt{2}} \la 1(a),2(b) \mid d(a,b)\ra\int \,d3\,d4\la 3,4\mid V_{\mathcal{A}}\mid 1,2\ra a^\dagger_{3}a^\dagger_{4}
\eol&&+\frac{1}{\sqrt{2}}\int d\,1 d\,2 \la 1(a),2(b) \mid d(a,b)\ra a_1^\dagger\int \,d3\,a^\dagger_{3}\mathcal{V}(3, 2).\label{VAcomm}\eeq
Using the eigenvalue equation (\ref{phdeq}) satisfied by $ \la 1(a),2(b) \mid d(a,b)\ra$ the first term on the right in eq.(\ref{VAcomm}) can be written
\beq \int d\,1 d\,2\, \la 1(a),2(b) \mid d(a,b)\ra\frac{1}{2}\int \,d3\,d4\la 3,4\mid V_{\mathcal{A}}\mid 1,2\ra a^\dagger_{3}a^\dagger_{4}
=(E_{d}-T_1-T_2) \la 1(a),2(b) \mid d(a,b)\ra a^\dagger_{1}a^\dagger_{2},\label{Vphid}\eeq
so that
\beq [H,A^\dagger_d]_-=E_{d}A^\dagger_{d}+\frac{1}{\sqrt{2}}\int d\,1 d\,2 \la 1(a),2(b) \mid d(a,b)\ra a^\dagger_{1}\int \,d3\,a^\dagger_{3}\mathcal{V}(3, 2),\label{HAcomm2}\eeq
or
\beq [H,A^\dagger_d]_-=E_{d}A^\dagger_d+J^\dagger_{d},\label{HAcomm3}\eeq
where
\beq J^\dagger_{d}=\int d1\,d2\,\frac{1}{\sqrt{2}} \la 1(a),2(b) \mid d(a,b)\ra a^\dagger_{1}\int \,d3\,a^\dagger_{3}\,\mathcal{V}(3, 2).\label{Jdagger}\eeq
This can also be written in terms of the commutator $[V,a_2^\dagger]_-$ using the result (\ref{Vadaggercomm}).
\beq J^\dagger_{d}=2\int d1\,d2\,\frac{1}{\sqrt{2}} \la 1(a),2(b) \mid d(a,b)\ra a^\dagger_{1}[V,a_2^\dagger]_-.\label{Jdagger2}\eeq

The Hermitian conjugate of eq.(\ref{HAcomm2}) gives
\beq [A_{d},H]_- = E_{K,d_0}A_d+J_d,\label{HAcomm4}\eeq
where
\beq J_{d}&&=\int d1\, \int \,d3\,\int\, d2\,\frac{1}{\sqrt{2}}(\frac{1}{\sqrt{2}} \la 1(a),2(b) \mid d(a,b)\ra)^*\mathcal{V}(2, 3) a_{3}a_{1}\eol &&=2\int d1\, d2\,\frac{1}{\sqrt{2}}( \la 1(a),2(b) \mid d(a,b)\ra)^*[a_2,V]_-a_1,\label{J}\eeq
and it has ben assumed that $V$ is Hermitian so that \cite{Johnson19} 
\beq \mathcal{V}(3, 2)^\dagger=\mathcal{V}(2, 3).\label{Vhermitian}\eeq

\section{Matrix element of a general non-local momentum conserving two-body interaction in a three-body space.}\label{gen2in3}
The general form of a momentum conserving non-local $b-A$  interaction in the space of bodies $a,b,A$ is 
\beq \la \ve{r}'_1(a), \ve{r}_3(b),\ve{r'}_A,n' \mid V_{bA}\mid\ve{r}_1(a), \ve{r}_2(b),\ve{r}_A,n \ra =\delta(\ve{r}_1-\ve{r}'_1)\delta(\ve{R}'_{bA}-\ve{R}_{bA}) \la \ve{r}'_{bA},n'\mid v_{bA}\mid\ve{r}_{bA},n\ra,\label{bAnonlocal}\eeq
where the various $\ve{r}$ are coordinates with respect to some arbitrary origin. $\ve{R}_{bA},\,\ve{R}'_{bA}$ are $b-A$ centre-of-mass coordinates defined by
\beq \ve{R}_{bA}=\frac{(\ve{r}_2+A\ve{r}_A)}{(A+1)},\,\,\, \ve{R}'_{bA}=\frac{(\ve{r}_3+A\ve{r}'_A)}{(A+1)}\label{RbA}\eeq
and
\beq \ve{r}_{bA}=(\ve{r}_2-\ve{r}_A),\,\,\,\ve{r}'_{bA}=(\ve{r}_3-\ve{r}'_A). \label{rbA}\eeq
To compare with the matrix element in eq.(\ref{UbarSP51}) it is useful to transform  (\ref{bAnonlocal}) to a mixed basis in which $a$ and $b$ have definite momentum, the position of $A$ is $\ve{r'}_A=0$ in the left-hand ket and in the right-hand ket $A$ has definite momentum $-(\ve{k}_1+\ve{k}_2)$ (as in the over-all c.m. system). The result is
\beq \la \ve{k}'_1(a), \ve{k}_3(b),\ve{r'}_A=0,n' \mid V_{bA}\mid\ve{k}_1(a), \ve{k}_2(b),-(\ve{k}_1+\ve{k}_2),n \ra =\delta(\ve{k}_1-\ve{k}'_1)  \la(\ve{k}_3+\frac{\ve{k}_1}{(A+1)}),n'\mid v_{bA} \mid (\ve{k}_2+\frac{\ve{k}_1}{(A+1)}),n\ra,\eol\label{bAnonlocal2}\eeq
where
\beq \la\ve{k}',n'\mid v_{bA}\mid\ve{k},n)\ra=\int\,d\ve{r}'_{bA}d\,\ve{r}_{bA}\frac{1}{(2\pi)^3}\exp (-\imath \ve{r}'_{bA}.\ve{k}' )\la \ve{r}'_{bA},n'\mid v_{bA}\mid\ve{r}_{bA},n\ra\exp (+\imath \ve{r}_{bA}.\ve{k} ). \label{vk1k2k3}\eeq
Eq.(\ref{bAnonlocal2}) shows that the dependence of the $v_{bA}$ matrix element on $\ve{k}_1$ disappears in the limit $A\rightarrow \infty$. For a local interaction
\beq \la \ve{r}'_{bA},n'\mid v_{bA}\mid\ve{r}_{bA},n\ra=\delta(\ve{r}'_{bA}-\ve{r}_{bA})v_{bA}(n',n,\ve{r}'_{bA}), \label{localVbA}\eeq
 and the matrix element $\la\ve{k}',n'\mid v_{bA}\mid\ve{k},n)\ra$ depends only on the momentum transfer $\ve{k}'-\ve{k}$. The dependence  on $\ve{k}_1$ then disappears for any $A$.  
\section{Overlap formalism and $d+A$ scattering and reaction theory.}\label{E}

     Because the states $A^\dagger_d \mid -\ve{K}_d, \psi_n\ra\ra$  do not form an orthogonal zero-momentum basis in Fock space,  the construction of the complete Fock space scattering state $\mid \Psi^{(\epsilon)}_{d_0,0}\ra\ra$ from the overlaps $\la\la \Psi(n,\ve{x}=0\mid A_{d}\mid \Psi^{\epsilon}_{d_0,0}\ra\ra$ requires the construction of the  analogue of the matrix defined in eq.(\ref{Kbarnkoko}) that provides the  similar step in the nucleon$+A$ case. However, this step is not necessary if all that is required is to calculate scattering crosssections. For example, in the nucleon$+A$ case  the transition matrix for elastic and inelastic scattering can be obtained from $\mid \Psi^{(\epsilon)}_{d_0,0}\ra\ra$ by using the exact expression \cite{Villars1967}
   \beq \langle \ve{k}_n,n\mid T(E) \mid \ve{k}_0,0 \rangle &=&- (2\pi)^{3/2}\la\la\Psi(n,\ve{x}=0)\mid[V,a_{\ve{k}_n}]\mid \Psi^{(+)}_{\ve{k}_0,0}\ra\ra.\eol &&\label{Tn02}\eeq
   The commutator $[V,a_{\ve{k}_n}]$ can be written \cite{Johnson19}, eq.(48),
   \beq [V,a_{\ve{k}_n}]=-\frac{1}{2}\int\,d\ve{k}_3\mathcal{V}(\ve{k}_n,\ve{k}_3)a_{\ve{k}_3},\label{aknV}\eeq
   where $\mathcal{V}$ is defined in eq.(\ref{Vkk12}). This result allows the overlap functions to be introduced in the right-hand-side of eq.(\ref{Tn02}) and these can be expressed in terms of the quantities $\la\la\Psi(n',\ve{x}=0 )\mid a_{\ve{k}_1}\mid \Psi_{\ve{k}_0,0}^{(\epsilon)}\ra\ra $ satisfying the. coupled equations (\ref{Beqcoupled2}). The matrix elements of $\mathcal{V}$ between $A$-nucleon target states can then be written in terms of the operator $\hat{\mathcal{U}}$ defined in eq.(\ref{Uint}). Eq. (\ref{Tn02}) becomes
   \beq \langle \ve{k}_n,n\mid T(E) \mid \ve{k}_0,0 \rangle &&=(2\pi)^{3/2}\int\,d\ve{k}'\sum_{n'}\la \ve{k}_n,n\mid\hat{\mathcal{U}}\mid \ve{k}',n'\ra\la\la\Psi(n',\ve{x}=0)\mid a_{\ve{k}'}\mid \Psi^{(+)}_{\ve{k}_0,0}\ra\ra,\label{Tn024}\eeq 
   or in terms of $B$-space quantities
   \beq \langle \ve{k}_n,n\mid T(E) \mid \ve{k}_0,0 \rangle &&=(2\pi)^{3/2}\int\,d\ve{k}'\sum_{n'}\la \ve{k}_n,n\mid\hat{\mathcal{U}}\mid \ve{k}',n'\ra\la\la\Psi(n',\ve{x}=0)\mid a_{\ve{k}'}\mid \Psi^{(+)}_{\ve{k}_0,0}\ra\ra,\label{Tn024}\eeq 
 This  result means that the transition amplitude, $S$-matrix and phase shifts in the elastic case  can be deduced by examining the asymptotic form of the solution of the coupled equations (\ref{dAscatt6}) or (\ref{Beq}) in the limit $\epsilon\rightarrow 0^+.$ Of course, care must be taken with the way this limit is taken. In the three-body case it may be preferable to convert these equations to the Faddeev or related forms where the limit can be taken straightforwardly. 
   
   \section{The $d$-space density matrix and the source term in the $B$-space coupled equations.}\label{F}   
   In this Section the inhomogeneous term in the coupled equations (\ref{inhomd}) is examined using techniques similar  to those used in Appendix \ref{B} for the nucleon$+A$ case. 
   
   In eq.(\ref{inhomd}) the inhomogeneous term has the form  
\beq  \la\la\Psi(n,\ve{x}=0)\mid A_{d} A^\dagger_{d_0}\mid -\ve{K_{d_0}}, \psi_{0}\ra\ra&&=\delta_{d,d_0}\delta_{n,0}\eol
&&-2\int d\ve{k}_2\, d\ve{k}_3  \rho_{d,d_0}(\ve{k}_3,\ve{k}_2) \la\la\Psi(n,\ve{x}=0)\mid a^\dagger_{\ve{k}_2}a_{\ve{k}_3}\mid -\ve{K_{d_0}}, \psi_{0}\ra\ra.\eol&&
 \label{inhomd2}\eeq
   
   The first term on the right gives rise to coupled equations with standard boundary conditions in the limit $\epsilon \rightarrow 0+$. The second term defines a contribution
   \beq \bar{\mathcal{K}}_{n,d;0,d_0}= 2\int d\ve{k}_2\, d\ve{k}_3  \rho_{d,d_0}(\ve{k}_3,\ve{k}_2)\la\la\Psi(n,\ve{x}=0)\mid a^\dagger_{\ve{k}_2}a_{\ve{k}_3}\mid -\ve{K_{d_0}},\psi_{0}\ra\ra \label{Kbard}\eeq
   and involves a more general set of target density matrix elements than in Appendix \ref{B} 
   
    The definition given in eq.(\ref{rhodd'}) is
  \beq \rho_{d,d'}(3,2)=\int\,d1\,\la d(a,b)\mid 1(a),3(b)\ra \la 1(a),2(b) \mid d'(a,b)\ra.\label{rhodd'2}\eeq
  With momentum space for the single nucleon basis and the definitions of Section \ref{General n-p} it is found that
  \beq \rho_{d,d'}(\ve{k}_3,\ve{k}_2)&&=\int\,d\ve{k}_1\,\la d(a,b)\mid \ve{k}_1,\ve{k}_3\ra \la \ve{k}_1(a),\ve{k}_2(b) \mid d'(a,b)\ra\eol
  &&=\int\,d\ve{k}_1\,\delta(\ve{k}_1+\ve{k}_3-\ve{K}_d)\phi^*_{d} ((\ve{k}_1-\ve{k}_3)/2)\eol&&\,\,\,\,\,\,\,\,\,\,\,\,\,\,\,\,\,\,\,\,\,\times \delta(\ve{k}_1+\ve{k}_2-\ve{K}_{d'}) \phi_{d'}((\ve{k}_1-\ve{k}_2)/2) \eol
  &&=\delta(\ve{k}_3-\ve{k}_2-(\ve{K}_d-\ve{K}_{d'}))\phi^*_{d}(\frac{\ve{K}_d}{2}-\ve{k}_3) \phi_{d'}( \frac{\ve{K}_{d'}}{2}-\ve{k}_2),\label{rhodd'3}\eeq
  where $\phi_d$ denotes the intrinsic components of either the deuteron ground state  or one of the continuum states introduced in eqs.(\ref{KS}) and (\ref{KS3}). For present purposes explicit references to nucleon intrinsic spin and isospin coordinates are omitted.
  
  In the notation of Section \ref{BdA}, the contribution of the $\bar{\mathcal{K}}$  term to the solution of eqs.(\ref{inhomd}) is determined by 
 \beq \frac{\imath \epsilon}{E+\imath \epsilon-\bar{\mathcal{H}}_0}\bar{\mathcal{K}}\mid 0,d_0\ra&&=\sum_{d,n}\mid d,n\ra \frac{\imath \epsilon}{E+\imath \epsilon-\frac{\hbar^2K_{d}^2}{4m}-\epsilon_{d}-E_{n}}\bar{\mathcal{K}}_{n,d;0,d_0}\eol
 &&=\sum_{d,n}\mid d,n\ra \frac{\imath \epsilon}{E+\imath \epsilon-\frac{\hbar^2K_{d}^2}{4m}-\epsilon_{d}-E_{n}}2\int d\ve{k}_2\, d\ve{k}_3  \delta(\ve{k}_3-\ve{k}_2-(\ve{K}_d-\ve{K}_{d_0}))\eol
 &&\times\phi^*_{d}(\frac{\ve{K}_d}{2}-\ve{k}_3) \phi_{d_0}( \frac{\ve{K}_{d_0}}{2}-\ve{k}_2)
 \la\la\Psi(n,\ve{x}=0)\mid a^\dagger_{\ve{k}_2}a_{\ve{k}_3}\mid -\ve{K_{d_0}}, \psi_{0}\ra\ra\eol&&\label{G0K}\eeq
A slightly more complicated calculation for $\la\la\Psi(n,\ve{x}=0)\mid a^\dagger_{\ve{k}_2}a_{\ve{k}_3}\mid -\ve{K_{d_0}}, \psi_{0}\ra\ra$ than in the nucleon$+A$ case gives the result
\beq\la\la\Psi(n,\ve{x}=0)\mid a^\dagger_{\ve{k}_2}a_{\ve{k}_3}\mid -\ve{K_{d_0}}, \psi_{0}\ra\ra&&\!\!\!\!\!\!=\frac{A}{(2\pi)^3}\int d\ve{\chi}_{1}\, \dots d\ve{\chi}_{(A-2)}d\ve{\chi}_{(A-1)}d\ve{\chi}'_{(A-1)} \exp(\imath (\frac{(A-1)\ve{k}_2+\ve{K}_{d_0}+\ve{k}_3}{A}).\ve{\chi}'_{(A-1)})\eol &&\times \psi^*_n(\ve{\chi}_1,\dots\,\ve{\chi}_{(A-2)},\ve{\chi}'_{(A-1)})\eol
&&\times \exp(-\imath (\ve{k}_3+\frac{\ve{K}_{d_0}}{A}).\ve{\chi}_{(A-1)})\psi_0(\ve{\chi}_1,\dots\,\ve{\chi}_{(A-2)},\ve{\chi}_{(A-1)}),\eol
 &&\!\!\!\!\!\!=\int d\ve{\chi}'_{(A-1)} \exp(\imath (\frac{(A-1)\ve{k}_2+\ve{K}_{d_0}+\ve{k}_3}{A}).\ve{\chi}'_{(A-1)})F_{n,0}((\ve{k}_3+\frac{\ve{K}_{d_0}}{A}),\ve{\chi}'_{(A-1)}),\eol && \label{DM125}\eeq
 where
 \beq F_{n,0}(\ve{q},\ve{\chi}'_{(A-1)})&&=\frac{A}{(2\pi)^3}\int d\ve{\chi}_{1}\, \dots d\ve{\chi}_{(A-1)} \psi^*_n(\ve{\chi}_1,\dots\,\ve{\chi}_{(A-2)},\ve{\chi}'_{(A-1)})\eol
&&\times \exp(-\imath \ve{q}.\ve{\chi}_{(A-1)})\psi_0(\ve{\chi}_1,\dots\,\ve{\chi}_{(A-2)},\ve{\chi}_{(A-1)})\label{Fn02}\eeq

Inserting this into eq.(\ref{inhomd3}) gives 
\beq \bar{\mathcal{K}}_{n,d;0,d_0}&&=2\int d\ve{k}_2\, d\ve{k}_3  \delta(\ve{k}_3-\ve{k}_2-(\ve{K}_d-\ve{K}_{d_0}))\phi^*_{d}(\frac{\ve{K}_d}{2}-\ve{k}_3) \phi_{d_0}( \frac{\ve{K}_{d_0}}{2}-\ve{k}_2)\eol&&\times  \int d\ve{\chi}'_{(A-1)} \exp(\imath (\frac{(A-1)\ve{k}_2+\ve{K}_{d_0}+\ve{k}_3}{A}).\ve{\chi}'_{(A-1)})F_{n,0}(\ve{k}_3+\frac{\ve{K}_{d_0}}{A},\ve{\chi}'_{(A-1)}),\eol&&
 \label{inhomd3}\eeq
which, putting $\ve{k}_2=\frac{\ve{K}_{d_0}}{2}-\ve{k}$, reduces to
\beq \bar{\mathcal{K}}_{n,d;0,d_0}&&=2\int d\ve{k}\,\phi^*_{d}(\frac{\ve{K}_{d_0}-\ve{K}_d}{2}+\ve{k}) \phi_{d_0}( \ve{k})\eol&&\times  \int d\ve{\chi}'_{(A-1)} \exp(\imath (\frac{\ve{K}_{d_0}}{2}+\frac{\ve{K}_{d}}{A}-\ve{k}).\ve{\chi}'_{(A-1)})F_{n,0}(\ve{K}_d+\frac{(2-A)\ve{K}_{d_0}}{2A}-\ve{k},\ve{\chi}'_{(A-1)}),\eol&&
 \label{inhomd4}\eeq
 
The result of using this expression in eq.(\ref{G0K}) is 
 \beq \frac{\imath \epsilon}{E+\imath \epsilon-\bar{\mathcal{H}}_0}\bar{\mathcal{K}}\mid 0,d_0\ra&&=\imath \epsilon\sum_{d,n}\mid d,n\ra \frac{1}{E+\imath \epsilon-\frac{(A+2)\hbar^2K_{d}^2}{4mA}-\epsilon_{d}-E_{n}}\bar{\mathcal{K}}_{n,d;0,d_0}.\eol
 &&\label{G0KAgeneral}\eeq
For bound deuteron and incident target states the quantity multiplying $\imath \epsilon $ on the right hand side will have a finite limit for $\epsilon \rightarrow 0$ except possibly if $\bar{\mathcal{K}}_{n,d;0,d_0}$ has a delta-function behaviour for some set of values of $n,d$ for which $E-\frac{(A+2)\hbar^2K_{d}^2}{4mA}-\epsilon_{d}-E_{n}$ vanishes. An examination of the explicit expression (\ref{inhomd4}) together with eq.(\ref{Fn02}) for the definition of $F_{n,0}$ shows that the most likely way this would occur is if $\phi_d$ and target state $\psi_n$ were in the continuum. This situation can be examined by modelling $\phi_d$ as a plane wave of momentum $\ve{k}_d$ and energy $\frac{\hbar^2k_d^2}{m}$ and the overlap 
\beq F_{n,0}(\ve{q},\ve{\chi}'_{(A-1)})=\int d\ve{\chi}_{1}\, \dots d\ve{\chi}_{(A-1)} \psi^*_n(\ve{\chi}_1,\dots\,\ve{\chi}_{(A-2)},\ve{\chi}'_{(A-1)})\exp(-\imath \ve{q}.\ve{\chi}_{(A-1)})\psi_0(\ve{\chi}_1,\dots\,\ve{\chi}_{(A-2)},\ve{\chi}_{(A-1)}),\label{n0overlap}\eeq
as a plane wave of momentum $\ve{k}_n$ in the variable $\ve{\chi}_{(A-1)}'$ and energy $\frac{A\hbar^2k_n^2}{2(A-1)m}$ satisfying ($A>1$)
\beq E_n=\frac{A\hbar^2k_n^2}{2(A-1)m}+E^{(A-1)}, \label{kn}\eeq
where $E^{(A-1)}$ is the energy of a state of the $(A-1)$ nucleon zero momentum state $\phi^{(A-1)}$ remaining after the disassociation of the $A$-nucleon state $n$.

With these definitions and ignoring irrelevant spin dependent quantities constants $\bar{\mathcal{K}}_{n,d;0,d_0}$ becomes
 \beq \bar{\mathcal{K}}_{n,d;0,d_0}&&=\int d\ve{k}\,\delta(\frac{\ve{K}_{d_0}-\ve{K}_d}{2}+\ve{k}-\ve{k}_d),\eol&&
 =\delta(\ve{K}_{d_0}+\frac{(2-A)\ve{K}_{d}}{2A}-\ve{k}_d-\ve{k}_n) \phi_{d_0}( \ve{k}_d-\frac{\ve{K}_{d_0}-\ve{K}_d}{2})  f_{n,0}(\frac{\ve{K}_d}{2}+\frac{\ve{K}_{d_0}}{A}-\ve{k}_d)
 \label{inhomd422}\eeq
 where
 \beq f_{n,0}(\frac{\ve{K}_d}{2}+\frac{\ve{K}_{d_0}}{A}-\ve{k}_d)&&=\int d\ve{\chi}_{1}\, \dots d\ve{\chi}_{(A-1)} \phi^{(A-1)}(\ve{\chi}_1,\dots\,\ve{\chi}_{(A-2)})^*\eol
&&\times \exp(-\imath (\frac{\ve{K}_d}{2}+\frac{\ve{K}_{d_0}}{A}-\ve{k}_d).\ve{\chi}_{(A-1)})\psi_0(\ve{\chi}_1,\dots\,\ve{\chi}_{(A-2)},\ve{\chi}_{(A-1)})    \label{fm0}\eeq
Inserting this expression into eq.(\ref{G0KAgeneral}) gives
\beq \frac{\imath \epsilon}{E+\imath \epsilon-\bar{\mathcal{H}}_0}\bar{\mathcal{K}}\mid 0,d_0\ra&&=\imath \epsilon\sum_{d,n}\mid d,n\ra \frac{1}{E+\imath \epsilon-\frac{(A+2)\hbar^2K_{d}^2}{4mA}-\epsilon_{d}-E_{n}}\bar{\mathcal{K}}_{n,d;0,d_0}\eol
 &&=\imath \epsilon\sum_{d,n}\mid d,n\ra \frac{1}{E+\imath \epsilon-\frac{(A+2)\hbar^2K_{d}^2}{4mA}-\epsilon_{d}-E_{n}}
 \eol&&\times \delta(\ve{K}_{d_0}+\frac{(2-A)\ve{K}_{d}}{2A}-\ve{k}_d-\ve{k}_n) \phi_{d_0}( \ve{k}_d-\frac{\ve{K}_{d_0}-\ve{K}_d}{2})  f_{n,0}(\frac{\ve{K}_d}{2}+\frac{\ve{K}_{d_0}}{A}-\ve{k}_d)\eol
 &&=\imath \epsilon\int d\ve{K}_d\sum_n\int d\ve{k}_n\mid d,n\ra \frac{1}{E+\imath \epsilon-\frac{(A+2)\hbar^2K_{d}^2}{4mA}-\frac{\hbar^2(\ve{K}_{d_0}+\frac{(2-A)\ve{K}_d}{2A}-\ve{k}_n)^2}{m}-\frac{A\hbar^2k_{n}^2}{2(A-1)m}-E_n^{(A-1)}}\eol&&\times\phi_{d_0}( \frac{\ve{K}_{d_0}}{2}+\frac{\ve{K}_{d}}{A}-\ve{k}_n)  f_{n,0}(\frac{(A-1)(\ve{K}_d-\ve{K}_{d_0})}{A}+\ve{k}_n).\label{G0KAgeneral2}\eeq
 If the deuteron ground state and the target ground state are bound the functions $\psi_0$ and $ f_{n,0}$ are non-singular functions of the integration variables $\ve{K}_d$ and $\ve{k}_n$ and therefore the coefficient of $\imath \epsilon$ on the right-hand-side of (\ref{G0KAgeneral2}) has a finite limit for $\epsilon\rightarrow 0$. It follows that
  \beq \lim_{\epsilon\rightarrow 0}\frac{\imath \epsilon}{E+\imath \epsilon-\bar{\mathcal{H}}_0}\bar{\mathcal{K}}\mid 0,d_0\ra&&=0.\label{G0KAgeneral3}\eeq
 \subsection{$A=1$ case.}
 Eq.(\ref{kn}) is clearly not applicable when $A=1$. For this special case eq.(\ref{G0K})can be evaluated by using
  \beq \mid  -\ve{K}_{d_0}, \psi_{0}\ra\ra &&\rightarrow (2\pi)^{(3/2)}a^\dagger_{-\ve{K}_{d_0},\sigma_{0}}\mid 0\ra\ra\eol
 \mid \Psi(n,\ve{x}=0)\ra&&\rightarrow \psi^\dagger(\ve{r}=0,\sigma_n)\mid 0\ra\ra. \label{A122}\eeq
 Note that when $A=1$ the index $n$ is just a set of spin and isospin eigenvalues denoted $\sigma_n$ for short. Using eq.(\ref{akpsidaggercomm2}) gives 
  \beq \la\la\Psi(n,\ve{x}=0)\mid a^\dagger_{\ve{k}_2}a_{\ve{k}_3}\mid  -\ve{K}_{d_0}, \psi_{0}\ra\ra\rightarrow \delta(\ve{k}_3+\ve{K}_{d_0})\delta_{\sigma_3,\sigma_{0}}\delta_{\sigma_2,\sigma_n},\label{A1222}\eeq
 and eq.(\ref{inhomd4}) is replaced by
 \beq \bar{\mathcal{K}}_{\sigma_n,d;\sigma_0,d_0}&&=2\int d\ve{k}_2\, d\ve{k}_3  \delta(\ve{k}_3-\ve{k}_2-(\ve{K}_d-\ve{K}_{d_0}))\phi^*_{d}(\frac{\ve{K}_d}{2}-\ve{k}_3) \phi_{d_0}( \frac{\ve{K}_{d_0}}{2}-\ve{k}_2) \delta(\ve{k}_3+\ve{K}_{d_0})\delta_{\sigma_3,\sigma_{0}}\delta_{\sigma_2,\sigma_n}
 \eol&&=2\int d\ve{k}_2\,   \delta(-\ve{k}_2-\ve{K}_d)\phi^*_{d}(\frac{\ve{K}_d}{2}+\ve{K}_{d_0}) \phi_{d_0}( \frac{\ve{K}_{d_0}}{2}-\ve{k}_2) \delta_{\sigma_3,\sigma_{0}}\delta_{\sigma_2,\sigma_n}\eol&&
 =2\phi^*_{d}(\frac{\ve{K}_d}{2}+\ve{K}_{d_0}) \phi_{d_0}( \frac{\ve{K}_{d_0}}{2}+\ve{K}_d).
 \label{inhomd32}\eeq 
 Reference to spin and isospin eigenvalues that are irrelevant for the present discussion have been omitted.
  
  The analogue of eq.(\ref{G0KAgeneral2}) is now 
  \beq \frac{\imath \epsilon}{E+\imath \epsilon-\bar{\mathcal{H}}_0}\bar{\mathcal{K}}\mid 0,d_0\ra&&=\sum_{d,n}\mid d,n\ra \frac{\imath \epsilon}{E+\imath \epsilon-\frac{3\hbar^2K_{d}^2}{4m}-\epsilon_{d}}\bar{\mathcal{K}}_{n,d;0,d_0}\eol
 &&=\int d\ve{K}_d \sum_d\mid \ve{K}_d,d\ra \frac{\imath \epsilon}{\frac{3\hbar^2K_{d_0}^2}{4m}-\epsilon_0+\imath \epsilon-\frac{3\hbar^2K_{d}^2}{4m}-\epsilon_{d}}2\phi^*_{d}(\frac{\ve{K}_d}{2}+\ve{K}_{d_0}) \phi_{d_0}( \frac{\ve{K}_{d_0}}{2}+\ve{K}_d)\eol &&
 .\label{G0KA1}\eeq
When  $d=d_0,\,\,\epsilon_d=-\epsilon_0$, both the states $\phi_{d_0}$ and $\phi_{d}$ are non-singular functions of their arguments and give      
  \beq \lim_{\epsilon\rightarrow 0} \frac{\imath \epsilon}{E+\imath \epsilon-\bar{\mathcal{H}}_0}\bar{\mathcal{K}}\mid 0,d_0\ra&&=0 .\label{G0KA2}\eeq

A continuum state  $\phi_d$ c asn be modelled as a plane wave with momentum $\ve{k}_d$ and energy $\epsilon_{d}=\hbar^2k_d^2/m $ so that (ignoring spin considerations)
 \beq \frac{\imath \epsilon}{E+\imath \epsilon-\bar{\mathcal{H}}_0}\bar{\mathcal{K}}\mid 0,d_0\ra&&=\sum_{d,n}\mid d,n\ra \frac{\imath \epsilon}{E+\imath \epsilon-\frac{3\hbar^2K_{d}^2}{4m}-\epsilon_{d}-E_{n}}\bar{\mathcal{K}}_{n,d;0,d_0}\eol
 &&=\int d\ve{K}_d d\ve{k}_d \mid\ve{K}_d,\ve{k}_d\ra\frac{\imath \epsilon}{E+\imath \epsilon-\frac{3\hbar^2K_{d}^2}{4m}-\frac{\hbar^2k_d^2}{m}}2\delta(\ve{k}_d-\frac{\ve{K}_d}{2}-\ve{K}_{d_0}) \phi_{d_0}( \frac{\ve{K}_{d_0}}{2}+\ve{K}_d)\eol &&
 =\int d\ve{K}_d \mid\ve{K}_d,\ve{k}_d=\frac{\ve{K}_d}{2}+\ve{K}_{d_0})\ra \frac{\imath \epsilon}{\frac{3\hbar^2K_{d_0}^2}{4m}-\epsilon_0+\imath \epsilon-\frac{3\hbar^2K_{d}^2}{4m}-\frac{\hbar^2(\frac{\ve{K}_d}{2}+\ve{K}_{d_0})^2}{m}} 2\phi_{d_0}( \frac{\ve{K}_{d_0}}{2}+\ve{K}_d)\eol
 &&=\int d\ve{K}_d \mid\ve{K}_d,\ve{k}_d=\frac{\ve{K}_d}{2}+\ve{K}_{d_0})\ra \frac{\imath \epsilon}{-\frac{\hbar^2K_{d_0}^2}{4m}-\epsilon_0+\imath \epsilon-\frac{\hbar^2K_{d}^2}{m}-\frac{\hbar^2\ve{K}_d.\ve{K}_{d_0})}{m}} 2\phi_{d_0}( \frac{\ve{K}_{d_0}}{2}+\ve{K}_d),\eol &&
=\int d\ve{K}_d \mid\ve{K}_d,\ve{k}_d=\frac{\ve{K}_d}{2}+\ve{K}_{d_0})\ra \frac{\imath \epsilon}{-\frac{\hbar^2(\frac{\ve{K}_{d_0}}{2}+\ve{K}_d)^2}{m}-\epsilon_0+\imath \epsilon} 2\phi_{d_0}( \frac{\ve{K}_{d_0}}{2}+\ve{K}_d).\eol&&\label{G0KA2}\eeq
For a stable deuteron ground state $\epsilon_0>0 $ and $\phi_{d_0}$ is square-integrable and denominator in the last line of (\ref{G0KA2}) never vanishes. Hence the expression on the right-hand-side of eq.(\ref{G0KA2}) vanishes in the limit $\epsilon \rightarrow 0$ and the contribution from the non-delta function term on the right-hand side of eq.(\ref{inhomd2}) to the solution of the coupled equations (\ref{inhomd}) vanishes for $A=1$ as well as $A>1$.
\end{document}